\documentclass[journal]{IEEEtran}
\usepackage{amsmath,amsfonts,amsthm}
\usepackage{mathtools}
\usepackage{algpseudocode}
\usepackage{algorithm}
\usepackage{array}
\usepackage[caption=false]{subfig}
\usepackage{textcomp}
\usepackage{stfloats}
\usepackage{url}
\usepackage{verbatim}
\usepackage{graphicx}
\usepackage{cite}
\usepackage[dvipsnames,table]{xcolor}
\expandafter\def\expandafter\normalsize\expandafter{%
    \normalsize
    \setlength\abovedisplayskip{4pt}
   \setlength\belowdisplayskip{4pt}
    \setlength\abovedisplayshortskip{5pt}
   \setlength\belowdisplayshortskip{5pt}
}

\usepackage{microtype}

\algnewcommand\algorithmicinput{\textbf{Output:}}
\algnewcommand\Output{\item[\algorithmicinput]}

\let\oldReturn\Return
\renewcommand{\Return}{\State\oldReturn}

\begin{document}

\bstctlcite{IEEEexample:BSTcontrol}
\def\bydef{:=}
\def\bba{{\mathbb{a}}}
\def\bbb{{\mathbb{b}}}
\def\bbc{{\mathbb{c}}}
\def\bbd{{\mathbb{d}}}
\def\bbee{{\mathbb{e}}}
\def\bbff{{\mathbb{f}}}
\def\bbg{{\mathbb{g}}}
\def\bbh{{\mathbb{h}}}
\def\bbi{{\mathbb{i}}}
\def\bbj{{\mathbb{j}}}
\def\bbk{{\mathbb{k}}}
\def\bbl{{\mathbb{l}}}
\def\bbm{{\mathbb{m}}}
\def\bbn{{\mathbb{n}}}
\def\bbo{{\mathbb{o}}}
\def\bbp{{\mathbb{p}}}
\def\bbq{{\mathbb{q}}}
\def\bbr{{\mathbb{r}}}
\def\bbs{{\mathbb{s}}}
\def\bbt{{\mathbb{t}}}
\def\bbu{{\mathbb{u}}}
\def\bbv{{\mathbb{v}}}
\def\bbw{{\mathbb{w}}}
\def\bbx{{\mathbb{x}}}
\def\bby{{\mathbb{y}}}
\def\bbz{{\mathbb{z}}}
\def\bb0{{\mathbb{0}}}
\def\bbone{{\mathbb{1}}}

\def\bydef{:=}
\def\ba{{\mathbf{a}}}
\def\bb{{\mathbf{b}}}
\def\bc{{\mathbf{c}}}
\def\bd{{\mathbf{d}}}
\def\bee{{\mathbf{e}}}
\def\bff{{\mathbf{f}}}
\def\bg{{\mathbf{g}}}
\def\bh{{\mathbf{h}}}
\def\bi{{\mathbf{i}}}
\def\bj{{\mathbf{j}}}
\def\bk{{\mathbf{k}}}
\def\bl{{\mathbf{l}}}
\def\bmm{{\mathbf{m}}} 
\def\bn{{\mathbf{n}}}
\def\bo{{\mathbf{o}}}
\def\bp{{\mathbf{p}}}
\def\bq{{\mathbf{q}}}
\def\br{{\mathbf{r}}}
\def\bs{{\mathbf{s}}}
\def\bt{{\mathbf{t}}}
\def\bu{{\mathbf{u}}}
\def\bv{{\mathbf{v}}}
\def\bw{{\mathbf{w}}}
\def\bx{{\mathbf{x}}}
\def\by{{\mathbf{y}}}
\def\bz{{\mathbf{z}}}
\def\bzero{{\mathbf{0}}}
\def\bone{{\mathbf{1}}}

\def\bA{{\mathbf{A}}}
\def\bB{{\mathbf{B}}}
\def\bC{{\mathbf{C}}}
\def\bD{{\mathbf{D}}}
\def\bE{{\mathbf{E}}}
\def\bF{{\mathbf{F}}}
\def\bG{{\mathbf{G}}}
\def\bH{{\mathbf{H}}}
\def\bI{{\mathbf{I}}}
\def\bJ{{\mathbf{J}}}
\def\bK{{\mathbf{K}}}
\def\bL{{\mathbf{L}}}
\def\bM{{\mathbf{M}}}
\def\bN{{\mathbf{N}}}
\def\bO{{\mathbf{O}}}
\def\bP{{\mathbf{P}}}
\def\bQ{{\mathbf{Q}}}
\def\bR{{\mathbf{R}}}
\def\bS{{\mathbf{S}}}
\def\bT{{\mathbf{T}}}
\def\bU{{\mathbf{U}}}
\def\bV{{\mathbf{V}}}
\def\bW{{\mathbf{W}}}
\def\bX{{\mathbf{X}}}
\def\bY{{\mathbf{Y}}}
\def\bZ{{\mathbf{Z}}}

\def\bbA{{\mathbb{A}}}
\def\bbB{{\mathbb{B}}}
\def\bbC{{\mathbb{C}}}
\def\bbD{{\mathbb{D}}}
\def\bbE{{\mathbb{E}}}
\def\bbF{{\mathbb{F}}}
\def\bbG{{\mathbb{G}}}
\def\bbH{{\mathbb{H}}}
\def\bbI{{\mathbb{I}}}
\def\bbJ{{\mathbb{J}}}
\def\bbK{{\mathbb{K}}}
\def\bbL{{\mathbb{L}}}
\def\bbM{{\mathbb{M}}}
\def\bbN{{\mathbb{N}}}
\def\bbO{{\mathbb{O}}}
\def\bbP{{\mathbb{P}}}
\def\bbQ{{\mathbb{Q}}}
\def\bbR{{\mathbb{R}}}
\def\bbS{{\mathbb{S}}}
\def\bbT{{\mathbb{T}}}
\def\bbU{{\mathbb{U}}}
\def\bbV{{\mathbb{V}}}
\def\bbW{{\mathbb{W}}}
\def\bbX{{\mathbb{X}}}
\def\bbY{{\mathbb{Y}}}
\def\bbZ{{\mathbb{Z}}}

\def\cA{\mathcal{A}}
\def\cB{\mathcal{B}}
\def\cC{\mathcal{C}}
\def\cD{\mathcal{D}}
\def\cE{\mathcal{E}}
\def\cF{\mathcal{F}}
\def\cG{\mathcal{G}}
\def\cH{\mathcal{H}}
\def\cI{\mathcal{I}}
\def\cJ{\mathcal{J}}
\def\cK{\mathcal{K}}
\def\cL{\mathcal{L}}
\def\cM{\mathcal{M}}
\def\cN{\mathcal{N}}
\def\cO{\mathcal{O}}
\def\cP{\mathcal{P}}
\def\cQ{\mathcal{Q}}
\def\cR{\mathcal{R}}
\def\cS{\mathcal{S}}
\def\cT{\mathcal{T}}
\def\cU{\mathcal{U}}
\def\cV{\mathcal{V}}
\def\cW{\mathcal{W}}
\def\cX{\mathcal{X}}
\def\cY{\mathcal{Y}}
\def\cZ{\mathcal{Z}}

\def\sfA{\mathsf{A}}
\def\sfB{\mathsf{B}}
\def\sfC{\mathsf{C}}
\def\sfD{\mathsf{D}}
\def\sfE{\mathsf{E}}
\def\sfF{\mathsf{F}}
\def\sfG{\mathsf{G}}
\def\sfH{\mathsf{H}}
\def\sfI{\mathsf{I}}
\def\sfJ{\mathsf{J}}
\def\sfK{\mathsf{K}}
\def\sfL{\mathsf{L}}
\def\sfM{\mathsf{M}}
\def\sfN{\mathsf{N}}
\def\sfO{\mathsf{O}}
\def\sfP{\mathsf{P}}
\def\sfQ{\mathsf{Q}}
\def\sfR{\mathsf{R}}
\def\sfS{\mathsf{S}}
\def\sfT{\mathsf{T}}
\def\sfU{\mathsf{U}}
\def\sfV{\mathsf{V}}
\def\sfW{\mathsf{W}}
\def\sfX{\mathsf{X}}
\def\sfY{\mathsf{Y}}
\def\sfZ{\mathsf{Z}}

\def\bsfA{\bm{\mathsf{A}}}
\def\bsfB{\bm{\mathsf{B}}}
\def\bsfC{\bm{\mathsf{C}}}
\def\bsfD{\bm{\mathsf{D}}}
\def\bsfE{\bm{\mathsf{E}}}
\def\bsfF{\bm{\mathsf{F}}}
\def\bsfG{\bm{\mathsf{G}}}
\def\bsfH{\bm{\mathsf{H}}}
\def\bsfI{\bm{\mathsf{I}}}
\def\bsfJ{\bm{\mathsf{J}}}
\def\bsfK{\bm{\mathsf{K}}}
\def\bsfL{\bm{\mathsf{L}}}
\def\bsfM{\bm{\mathsf{M}}}
\def\bsfN{\bm{\mathsf{N}}}
\def\bsfO{\bm{\mathsf{O}}}
\def\bsfP{\bm{\mathsf{P}}}
\def\bsfQ{\bm{\mathsf{Q}}}
\def\bsfR{\bm{\mathsf{R}}}
\def\bsfS{\bm{\mathsf{S}}}
\def\bsfT{\bm{\mathsf{T}}}
\def\bsfU{\bm{\mathsf{U}}}
\def\bsfV{\bm{\mathsf{V}}}
\def\bsfW{\bm{\mathsf{W}}}
\def\bsfX{\bm{\mathsf{X}}}
\def\bsfY{\bm{\mathsf{Y}}}
\def\bsfZ{\bm{\mathsf{Z}}}

\def\bydef{:=}
\def\sfa{{\mathsf{a}}}
\def\sfb{{\mathsf{b}}}
\def\sfc{{\mathsf{c}}}
\def\sfd{{\mathsf{d}}}
\def\sfee{{\mathsf{e}}}
\def\sfff{{\mathsf{f}}}
\def\sfg{{\mathsf{g}}}
\def\sfh{{\mathsf{h}}}
\def\sfi{{\mathsf{i}}}
\def\sfj{{\mathsf{j}}}
\def\sfk{{\mathsf{k}}}
\def\sfl{{\mathsf{l}}}
\def\sfm{{\mathsf{m}}}
\def\sfn{{\mathsf{n}}}
\def\sfo{{\mathsf{o}}}
\def\sfp{{\mathsf{p}}}
\def\sfq{{\mathsf{q}}}
\def\sfr{{\mathsf{r}}}
\def\sfs{{\mathsf{s}}}
\def\sft{{\mathsf{t}}}
\def\sfu{{\mathsf{u}}}
\def\sfv{{\mathsf{v}}}
\def\sfw{{\mathsf{w}}}
\def\sfx{{\mathsf{x}}}
\def\sfy{{\mathsf{y}}}
\def\sfz{{\mathsf{z}}}
\def\sf0{{\mathsf{0}}}

\def\bsfa{{\bm{\mathsf{a}}}}
\def\bsfb{{\bm{\mathsf{b}}}}
\def\bsfc{{\bm{\mathsf{c}}}}
\def\bsfd{{\bm{\mathsf{d}}}}
\def\bsfee{{\bm{\mathsf{e}}}}
\def\bsfff{{\bm{\mathsf{f}}}}
\def\bsfg{{\bm{\mathsf{g}}}}
\def\bsfh{{\bm{\mathsf{h}}}}
\def\bsfi{{\bm{\mathsf{i}}}}
\def\bsfj{{\bm{\mathsf{j}}}}
\def\bsfk{{\bm{\mathsf{k}}}}
\def\bsfl{{\bm{\mathsf{l}}}}
\def\bsfm{{\bm{\mathsf{m}}}}
\def\bsfn{{\bm{\mathsf{n}}}}
\def\bsfo{{\bm{\mathsf{o}}}}
\def\bsfp{{\bm{\mathsf{p}}}}
\def\bsfq{{\bm{\mathsf{q}}}}
\def\bsfr{{\bm{\mathsf{r}}}}
\def\bsfs{{\bm{\mathsf{s}}}}
\def\bsft{{\bm{\mathsf{t}}}}
\def\bsfu{{\bm{\mathsf{u}}}}
\def\bsfv{{\bm{\mathsf{v}}}}
\def\bsfw{{\bm{\mathsf{w}}}}
\def\bsfx{{\bm{\mathsf{x}}}}
\def\bsfy{{\bm{\mathsf{y}}}}
\def\bsfz{{\bm{\mathsf{z}}}}
\def\bsf0{{\bm{\mathsf{0}}}}

\newcommand{\ngp}[1]{\textcolor{violet}{XXX Nuria - #1}}
\newcommand{\mb}[1]{\textcolor{orange}{XXX Murat - #1}}
\newcommand{\del}[1]{\textcolor{red}{#1}}
\newcommand{\add}[1]{\textcolor{blue}{#1}}

\newcommand{\Exp}[1]{{\mathbb{E}\left\{#1\right\}}}     
\newcommand{\diag}[1]{{\mathrm{diag}\left\{#1\right\}}} 
\newcommand{\trace}[1]{{\mathrm{Tr}\left\{#1\right\}}}  
\newcommand{\norm}[1]{\left\lVert#1\right\rVert}        

\newcommand{\rmF}{\mathrm{F}}               
\newcommand{\rmH}{\mathrm{H}}               
\newcommand{\rmT}{\mathrm{T}}               

\newcommand{\NBST}{N_{\mathrm{BS,T}}}       
\newcommand{\NBSR}{N_{\mathrm{BS,R}}}       
\newcommand{\NDL}{N_{\mathrm{DL}}}        
\newcommand{\NUL}{N_{\mathrm{UL}}}        

\newcommand{\NUE}{N_{\mathrm{UE}}}        

\newcommand{\NBSRF}{N_{\mathrm{BS,RF}}}     
\newcommand{\nRF}{n_{\mathrm{RF}}}          
\newcommand{\NDLRF}{N_{\mathrm{DL,RF}}}     
\newcommand{\NULRF}{N_{\mathrm{UL,RF}}}     

\newcommand{\NUERF}{N_{\mathrm{UE,RF}}}     

\newcommand{\NDLs}{N_{\rm DL,s}}            
\newcommand{\NULs}{N_{\rm UL,s}}            
\newcommand{\ns}{n_{\rm s}}                 

\newcommand{\yUL}{\mathbf{y}_{m,n}^{(j)}}   
\newcommand{\yBS}{\mathbf{y}_{m,n}}         
\newcommand{\yDL}{\mathbf{r}_{m,n}^{(i)}}   

\newcommand{\yBSrad}{y_{m,n}^{\rm rad}}         

\newcommand{\nUL}{\mathbf{n}_{m,n}}         
\newcommand{\nULo}{\mathbf{n}_{m,n}^{0}}         
\newcommand{\bnULo}{\bar{\mathbf{n}}_{m,n}^{0}}         
\newcommand{\bnULom}{\bar{\mathbf{n}}_{m}^{0}}         
\newcommand{\nULj}{\tilde{\mathbf{n}}_{m,n}^{(j)}}         
\newcommand{\nDL}{\mathbf{z}_{m,n}^{(i)}}   
\newcommand{\nDLi}{\tilde{\mathbf{z}}_{m,n}^{(i)}}   

\newcommand{\PtDL}{P_{\rm DL}}              
\newcommand{\PtUL}{P_{\rm UL}^{(j)}}              
\newcommand{\varDL}{\sigma_{i}^{2}}      
\newcommand{\varUL}{\sigma^{2}}      

\newcommand{\WRF}{\mathbf{W}_{\mathrm{RF}}}             
\newcommand{\WRFcom}{\mathbf{W}_{\mathrm{RF}}^{\rm com}}             
\newcommand{\WRFrad}{\mathbf{W}_{\mathrm{RF}}^{\rm rad}}             
\newcommand{\WBB}{\mathbf{W}_{\mathrm{BB},m}^{(j)}}     
\newcommand{\FBS}{\mathbf{F}_{m}^{(i)}}                 
\newcommand{\FBScom}{\mathbf{F}_{\mathrm{com},m}^{(i)}}                 
\newcommand{\FBSrad}{\mathbf{F}_{\mathrm{rad},m}^{(i)}}                 
\newcommand{\fBSrad}{\bff_{\mathrm{rad},m}^{(i)}}                 
\newcommand{\FBSi}{\mathbf{F}_{m}^{(i')}}               
\newcommand{\FBSRF}{\mathbf{F}_{\mathrm{RF}}}           
\newcommand{\FBSBB}{\mathbf{F}_{\mathrm{BB},m}^{(i)}}   
\newcommand{\FBSBBi}{\mathbf{F}_{\mathrm{BB},m}^{(i')}} 

\newcommand{\WDL}{\mathbf{U}_{m}^{(i)}}                 
\newcommand{\WDLi}{\mathbf{U}_{m}^{(i')}}                 
\newcommand{\WDLRF}{\mathbf{U}_{{\rm RF}}^{(i)}}            
\newcommand{\WDLBB}{\mathbf{U}_{{\rm BB},m}^{(i)}}            

\newcommand{\FUL}{\mathbf{V}_{m}^{(j)}}                 
\newcommand{\FULj}{\mathbf{V}_{m}^{(j')}}               
\newcommand{\FULRF}{\mathbf{V}_{{\rm RF}}^{(j)}}          
\newcommand{\FULRFj}{\mathbf{V}_{{\rm RF}}^{(j')}}        
\newcommand{\FULBB}{\mathbf{V}_{{\rm BB},m}^{(j)}}      
\newcommand{\FULBBj}{\mathbf{V}_{{\rm BB},m}^{(j')}}    

\newcommand{\HDL}{\mathbf{H}_{m}^{(i)}}                 
\newcommand{\HDLi}{\mathbf{H}_{m}^{(i')}}                 
\newcommand{\HUL}{\mathbf{G}_{m}^{(j)}}                 
\newcommand{\HULj}{\mathbf{G}_{m}^{(j')}}               
\newcommand{\Ht}{\mathbf{A}_{m,n}}                      
\newcommand{\HSI}{\mathbf{H}_{\rm SI}}                  

\newcommand{\sDL}{\mathbf{d}^{(i)}_{m,n}}
\newcommand{\sDLi}{\mathbf{d}^{(i')}_{m,n}}
\newcommand{\sUL}{\mathbf{s}^{(j)}_{m,n}}
\newcommand{\sULj}{\mathbf{s}^{(j')}_{m,n}}

\newcommand{\fc}{f_{\rm c}}                         
\newcommand{\deltaf}{\Delta f}                      
\newcommand{\Tcp}{T_{\rm cp}}                       
\newcommand{\Ts}{T_{\rm s}}                         

\newcommand{\alphaDL}{\alpha_{l}^{(i)}}             
\newcommand{\alphaUL}{\bar{\alpha}_{l}^{(j)}}       
\newcommand{\betat}{\beta_{k}}                      
\newcommand{\bbetat}{\bar{\beta}_{k}}                      

\newcommand{\aBST}{\mathbf{a}_{\rm BS,T}}           
\newcommand{\aBSR}{\mathbf{a}_{\rm BS,R}}           
\newcommand{\aDL}{\mathbf{a}_{{\rm DL}}}            
\newcommand{\aUL}{\mathbf{a}_{{\rm UL}}}            

\newcommand{\aUE}{\mathbf{a}_{{\rm UE}}}            

\newcommand{\LDL}{L_{\rm{DL}}^{(i)}}                
\newcommand{\LUL}{L_{\rm{UL}}^{(j)}}                

\newcommand{\UDL}{U_{{\rm{DL}}}}                    
\newcommand{\UUL}{U_{{\rm{UL}}}}                    

\newcommand{\thetaDL}{\vartheta_{l}^{(i)}}          
\newcommand{\phiDL}{\varphi_{l}^{(i)}}              

\newcommand{\thetaUL}{\bar{\vartheta}_{l}^{(j)}}    
\newcommand{\phiUL}{\bar{\varphi}_{l}^{(j)}}        

\newcommand{\tauDL}{\tau_{l}^{(i)}}                 
\newcommand{\tauUL}{\bar{\tau}_{l}^{(j)}}           

\newcommand{\thetat}{\theta_{k}}                    
\newcommand{\thetar}{\theta_{\rm r}}                
\newcommand{\taut}{\tau_{k}}                        
\newcommand{\fDk}{f_{\mathrm{D},k}}                 

\newcommand{\rateDL}{\cR_{m}^{(i)}}                 
\newcommand{\rateUL}{\bar{\cR}_{m}^{(j)}}           
\newcommand{\RnDLi}{\bR_{\tilde{\bz}_{m}}^{(i)}}                    
\newcommand{\RnULj}{\bR_{\tilde{\bn}_{m}}^{(j)}}                    

\newcommand{\SLNR}{\mathsf{SLNR}_{m}^{(i)}}         
\newcommand{\SLNRrad}{\overline{\mathsf{SLNR}}_{m}^{(i)}}         

\newcommand{\Bcov}{\bB_{m}^{(i)}}                   
\newcommand{\Bcovi}{\bB_{m}^{(i')}}                 
\newcommand{\bBcov}{\bar{\bB}_{m}^{(i)}}                   
\newcommand{\bBcovi}{\bar{\bB}_{m}^{(i')}}                 

\newcommand{\SINR}{\mathsf{SINR}_{\nRF}}            

\newcommand{\GT}{G_{\mathrm{T},m,\ns}^{(i)}}              
\newcommand{\GR}{G_{\mathrm{R},\nRF}}               
\newcommand{\Gcom}{G_{\mathrm{com},\nRF}}               
\newcommand{\GRloss}{G_{\mathrm{R},\nRF}^{\rm loss}}               
\newcommand{\Gcomloss}{G_{\mathrm{com},\nRF}^{\rm loss}}               

\newcommand{\tauT}{\tau_{\rm T}}                    
\newcommand{\tauR}{\tau_{\rm R}}                    
\newcommand{\taucom}{\tau_{\rm com}}                    

\title{Integrated Monostatic Sensing and Full-Duplex Multiuser Communication for mmWave Systems}

\author{Murat Bayraktar,~\IEEEmembership{Graduate Student Member,~IEEE,} Nuria González-Prelcic,~\IEEEmembership{Senior Member,~IEEE,} Mikko Valkama,~\IEEEmembership{Fellow,~IEEE,} Hao Chen, and Charlie Jianzhong Zhang,~\IEEEmembership{Fellow,~IEEE}
\thanks{N. Gonz\'{alez}-Prelcic and M. Bayraktar are with the
Department of Electrical and Computer Engineering, University of California San Diego, USA (e-mail: \{mbayraktar, ngprelcic\}@ucsd.edu).
M. Valkama is with the Unit of Electrical Engineering, Tampere University, Finland (e-mail: mikko.valkama@tuni.fi).
H. Chen and C. J. Zhang are with Samsung Research America, USA (e-mail: \{hao.chen1, jianzhong.z\}@samsung.com).}
}



\maketitle

\begin{abstract}
In this paper, we propose a hybrid precoding/combining framework for communication-centric integrated sensing and full-duplex (FD) communication operating at mmWave bands. The designed precoders and combiners enable multiuser (MU) FD communication while simultaneously supporting monostatic sensing in a frequency-selective setting. The joint design of precoders and combiners involves the mitigation of self-interference (SI) caused by simultaneous transmission and reception at the FD base station (BS). Additionally, MU interference needs to be handled by the precoder/combiner design. The resulting optimization problem involves non-convex constraints since hybrid analog/digital architectures utilize networks of phase shifters. To solve the proposed problem, we separate the optimization of each precoder/combiner, and design each one of them while fixing the others. The precoders at the FD BS are designed by reformulating the communication and sensing constraints as signal-to-leakage-plus-noise ratio (SLNR) maximization problems that consider SI and MU interference as leakage. Furthermore, we design the frequency-flat analog combiner such that the residual SI at the FD BS is minimized under communication and sensing gain constraints. Finally, we design an interference-aware digital combining stage that separates MU signals and target reflections. The communication performance and sensing results show that the proposed framework efficiently supports both functionalities simultaneously.
\end{abstract}


\section{Introduction}


\IEEEPARstart{I}{ntegrated} sensing and communication (ISAC) is one of the disruptive technologies to be incorporated into future wireless networks \cite{Liu2022JSAC,Proceedings2024NGP}. 
For such systems, it is possible to exploit a single waveform for both sensing and communication purposes. 
Orthogonal frequency division multiplexing (OFDM) is the most notable communication-centric transmission technique that is also utilized for sensing \cite{Barneto2019TMTT}. 
The integration of sensing and communication is especially interesting at mmWave bands, due to the operation with large arrays and bandwidths that enable high accuracy angle and delay estimation \cite{Heath2016JSTSP}. One of the potential network sensing modes in the context of ISAC systems is monostatic sensing 
enabled by a full duplex (FD) transceiver  \cite{Barneto2021WC,Proceedings2024NGP}.
Operation with a FD mmWave system introduces challenges associated to the mitigation of the self-interference (SI) while supporting communication and sensing \cite{Smida2024Proc}. 
The SI mitigation constraint complicates precoder/combiner design as SI creates a coupling between the transmitter (TX) and receiver (RX) sides of the FD transceiver. 
Although most works in the ISAC literature neglect this phenomenon by assuming perfect isolation between the TX and RX arrays \cite{Liu2022JSAC, Keskin2021TSP, Jiang2022SystJ, Bazzi2023TWC}, it is an unavoidable constraint that needs to be considered in the design of the physical layer of the FD system.

\subsection{Prior Work}

FD communication at mmWave bands has been extensively studied in the recent literature \cite{Roberts2021WC}. 
The main challenge in these systems is to reduce the SI before the analog-to-digital converters (ADCs) and low-noise amplifiers (LNAs) of the RX side of the FD transceiver, to prevent saturation and clipping. 
The SI levels can be $100$dB stronger than the intended received signal \cite{Smida2024Proc,Sabharwal2014JSAC}.
Although there are techniques to reduce SI by using analog cancellation methods, their realization at mmWave bands is costly due to the large number of antennas and the utilization of a hybrid analog/digital architecture \cite{Barneto2021WC,Roberts2021WC,Smida2024Proc}.
Because of this, precoders and combiners are also exploited for SI suppression at mmWave. However, the analog and hybrid analog/digital architectures employed at mmWave frequencies makes the design of precoders/combiners a challenging problem. 
We review a set of selected examples that consider this problem \cite{Valcarce2019FD,Satyanarayana2019TVT,Palacios2019GLOBECOM,Valcarce2020FD,Roberts2021TWC,Koc2021OJCOMS,Sheemar2022OJCOMS,Valcarce2022TWC}. 
In \cite{Valcarce2019FD}, the authors proposed the first SI-aware analog beamforming design for point-to-point FD mmWave systems. The proposed design utilizes an iterative procedure such that the entries of the analog beamformers are projected to the subspace that is orthogonal to the SI terms, which is followed by normalization to satisfy the unit-modulus constraint.
In \cite{Satyanarayana2019TVT}, the first hybrid architecture for FD mmWave multiple-input multiple-output (MIMO) systems was proposed. The authors first find the fully digital precoders/combiners that eliminate the SI while maximizing the sum rate, then they utilize a hybrid factorization algorithm. However, this approach does not guarantee that the residual SI at the output of the ADCs would be low.
Iterative hybrid precodering/combining solutions were developed by forcing the SI to zero at each precoder and combiner design stage for FD mmWave MIMO systems in \cite{Palacios2019GLOBECOM,Valcarce2020FD}. These architectures were successful in cancelling the SI due to the zero-forcing constraints.
In \cite{Roberts2021TWC}, the authors designed the hybrid precoders/combiners such that the residual SI is in the limited dynamic range of the LNAs and ADCs, instead of designing them in the orthogonal subspace of the SI.
In \cite{Koc2021OJCOMS}, the authors proposed a hybrid precoding/combining architecture that exploits slowly varying angular support of the channels to design the analog beamformers while suppressing the SI.
Hybrid precoders and combiners were designed for multiuser (MU) FD communication at mmWave in \cite{Sheemar2022OJCOMS}, where both SI and the noise caused by non-ideal hardware is suppressed while maximizing the weighted sum-rate. 
Finally, hybrid precoders and combiners were developed for FD MIMO systems with quantized phase-shifters in \cite{Valcarce2022TWC}.

Hybrid precoder and combiner design for ISAC systems that exploit a FD mmWave architecture is still in its infancy. 
Most of the initial work on this subject investigates the hybrid precoding/combining with only downlink (DL) communication, while the RX node of the FD base station (BS) is utilized for monostatic sensing \cite{Liyanaarachchi2021JCS,Barneto2022TCOM,Islam2022ICC,Islam2022GLOBECOM,Bayraktar2023CAMSAP,Bayraktar2024ICC}. In other words, there is no simultaneous DL and UL communication.
The works in \cite{Liyanaarachchi2021JCS,Barneto2022TCOM} consider multiple DL user equipments (UEs) with only a line-of-sight (LoS) path, and design the precoders at the FD BS by guaranteeing DL communication while achieving a high TX gain at the target angles. 
In general, the analog combiner at the FD BS is designed to maximize the RX gain at the target angle in the null-space of the SI. 
However, this solution cannot be realized with commonly used phase shifting networks, since the entries of the optimal combiner do not satisfy the unit modulus constraint. 
A hybrid beamforming framework is proposed for a similar setup with only one DL UE in \cite{Islam2022ICC} and multiple DL UEs in \cite{Islam2022GLOBECOM}. 
The authors utilize a hardware-based analog SI cancellation method, with residual SI being cancelled by the digital precoders at the FD BS. 
However, these works assume that the UE paths are a subset of the targets, which is not necessarily true for ISAC systems. 
Furthermore, they exploit DFT codeboks for the design of the analog precoders/combiners, limiting the achievable communication and sensing performance. 
To overcome this limitation, in our previous work we developed an SI-aware analog beam codebook design that is suitable for both communication and sensing purposes \cite{Bayraktar2023CAMSAP}. 
In another work \cite{Bayraktar2024ICC}, we proposed a hybrid precoding architecture to support simultaneous monostatic sensing and DL communication in a single-user scenario while suppressing the SI. Unlike previous work, we considered DL channels with multiple paths which are not necessarily reflected by a subset of targets. Furthermore, this architecture includes an analog combiner design that further reduces the residual SI while providing high RX gain. 

Supporting only DL communication for FD ISAC systems is inefficient since the power allocated for sensing at the precoding stage reduces the DL communication rate and the RX side of the FD BS is only utilized for sensing. 
Therefore, enabling simultaneous FD communication and sensing is a necessary goal to fully exploit FD ISAC systems. 
There are only a few initial works on integrated monostatic sensing and FD communication \cite{Liu2023JSAC,He2023JSAC,Talha2023Asilomar}. 
An integrated sensing and FD communication system exploiting a fully digital MIMO architecture that only considers the residual SI is studied in \cite{Liu2023JSAC,He2023JSAC}. 
These solutions are not applicable, however, at mmWave bands since hybrid analog/digital architecture needs to be considered at these frequencies, and the residual SI needs to be significantly lowered at the output of the analog combiners. The work in \cite{Talha2023Asilomar} proposes a hybrid precoding and combining architecture with a DL and an uplink (UL) UE.
However, the channels are assumed to have only LoS path, and the UEs are considered as targets. Additionally, DFT codebooks are utilized for analog precoder/combiner design. These assumptions simplify both the system model and the design problem. 

\subsection{Contributions}
In this paper, we tackle the challenging problem of enabling simultaneous monostatic sensing and FD MU communication for wideband mmWave MIMO systems that employ hybrid analog/digital architecture. To the best of our knowledge, this problem has not been solved before, not even for a simplistic narrowband mmWave MIMO system. One  challenging aspect of this problem is that MU interference and SI needs to be suppressed with the hybrid precoders/combiners which simultaneously enable communication and sensing. Specifically, the residual SI power should be lowered at the output of the analog combiner of the FD BS. Considering the unit modulus constraint due to the analog phase shifting network, this is a uniquely challenging problem. Our design aims to  jointly suppress the SI with the hybrid precoders and the analog combiner at the FD BS. The excessive number of constraints and also the coupling between the variables make the problem highly intractable. Thus, we consider an alternating optimization approach and optimize each precoder/combiner while keeping the others constant. This approach enables us to apply various optimization techniques to solve each subproblem efficiently. Furthermore, based on the proposed analog combiner and the digital radar processing architecture, we refine the initial angle estimates and employ an OFDM radar technique to estimate the range and velocity of the targets.
Our contributions can be summarized as follows:
\begin{itemize}
    \item We first formulate the optimization of the fully digital precoder at the FD BS that aims to: (i) maximize the sum spectral efficiency for DL UEs, (ii) provide high TX radar gains, and (iii) suppress SI. The maximization of sum spectral efficiency is reformulated as signal-to-leakage-plus-noise ratio (SLNR) maximization, where the leakage terms include MU interference and SI. Similarly, an SLNR maximization problem is formulated for sensing beam design. The generalized eigenvalue solution is employed for finding communication and sensing precoders, while their weighted sum is used as the final precoder. Then, a hybrid factorization algorithm is utilized to find the digital precoders, and user-independent and frequency-flat analog precoder. 
    \item Considering that the hybrid factorization degrades the SI suppression capability of the precoders, we formulate the analog combiner design as the minimization of the residual SI with some communication and sensing guarantees. To that end, we constrain the analog precoder to have high gains for sensing and UL communication beams. Since the prior work does not consider the reception of separate UL signals and target reflections in an SI-limited analog architecture, the proposed optimization problem is the first of its kind. The proposed analog combiner design problem is still non-convex. We circumvent this issue by using a convex relaxation technique complemented with a random block coordinate descent approach.
    \item The analog combiner operates as a spatial mask that enhances sensing and UL signals while suppressing the SI. 
    Thus, we propose an interference-aware digital processing stage.  
    To that end, we design digital combiners to isolate the signals of UL UEs. We also utilize minimum variance distortionless response (MVDR) beamformer at the output of the analog combiner to suppress the UL signals and enhance the target signals.
    \item The simulation results reveal that the proposed iterative combiner design converges to reasonably low residual SI levels. Furthermore, communication and sensing performance results demonstrate the effectiveness of the proposed design to simultaneously support sensing and FD MU communication at mmWave. Lastly, sensing results show that the interference caused by UL signals can be jointly mitigated by MVDR beamformer and OFDM radar.
\end{itemize}

\subsubsection*{Notation}
The following notation is used throughout the paper: 
Lowercase $x$ denotes a scalar, bold lowercase $\bx$ denotes a column vector, bold uppercase $\bX$ denotes a matrix, and $\cX$ represents a set. 
Moreover, the transpose and conjugate transpose of the matrix $\bX$ are denoted by $\bX^{\rm T}$ and $\bX^{\rm H}$, respectively. 
The Euclidean and Frobenius norms of the vector $\bx$ and the matrix $\bX$ are represented by $\norm{\bx}$ and $\norm{\bX}_{\rmF}$, respectively. 
The expectation and trace operators are denoted by $\Exp{\cdot}$ and $\trace{\cdot}$, respectively. 
The $p$-th element of the vector $\bx$ is denoted by $[\bx]_{p}$, and the $(p,q)$-th entry of the matrix $\bX$ is represented by $[\bX]_{p,q}$. 
Furthermore, $[\bX]_{p,:}$ and $[\bX]_{:,q}$ denote the $p$-th row and the $q$-th column of the matrix $\bX$, respectively. 
The identity matrix is represented by $\bI$, and the matrix with all zeros is denoted by $\mathbf{0}$. 

\section{System Model}

\begin{figure*}[!t]
    \centering
    \includegraphics[width=0.8\linewidth]{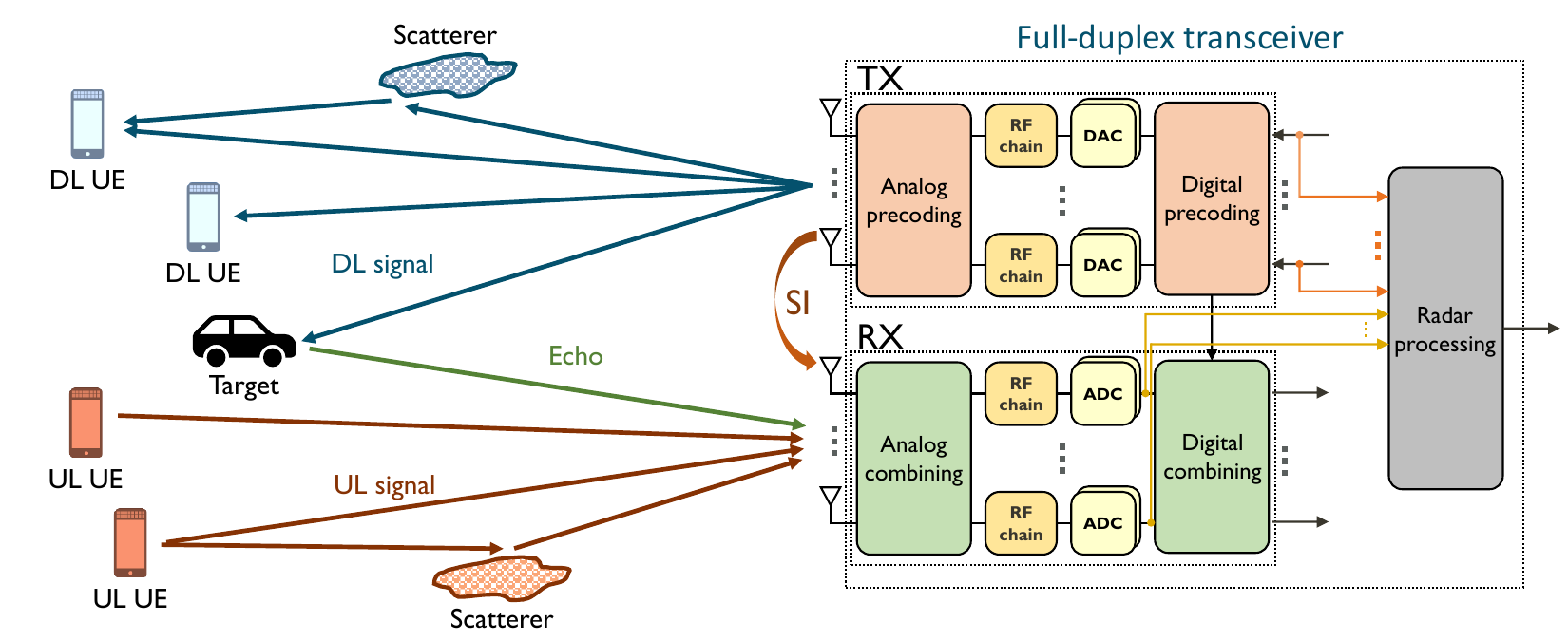}
    \caption{Illustration of the integrated monostatic sensing and FD MU communication system operating at mmWave bands.}
    \label{fig:illustration}
    \vspace{-10pt}
\end{figure*}

We consider a communication-centric ISAC system where the TX side of the FD BS sends data streams to multiple DL UEs, while the RX side simultaneously captures the reflections from targets and the UL data streams transmitted by multiple UEs, as illustrated in Fig~\ref{fig:illustration}.
We assume that perfect knowledge of the MU communication channel is available, in addition to  a coarse estimate of the target angles obtained in the previous frame.
In other words, we consider a tracking scenario for the radar and communication operation such that the angle, range and velocity of the targets have to be re-estimated, and the precoders/combiners have to be optimized for sensing and communication after the communication channel has been re-estimated.
Although there are multiple targets in the environment, we track a single target during a frame as in \cite{Liu2023JSAC,He2023JSAC}, while other targets are tracked in proceeding frames. Since the frame duration is very short at mmWave bands, this approach does not introduce any practical limitation for sensing. Consequently, parameters of all the targets can be estimated/updated in a reasonable duration. Furthermore, the targets can be tracked in a cyclic manner such that after estimating the parameters of all the targets in consecutive frames, the system can proceed updating the parameters of the targets one-by-one in the next frame. 

The system exploits OFDM modulation with $M$ subcarriers for signal transmission. The number of OFDM symbols in a frame is denoted by $N$. The number of antennas at the collocated TX and RX arrays of the FD BS are denoted by $\NBST$ and $\NBSR$, respectively. The FD BS is also equipped with $\NBSRF$ RF chains at both TX and RX ends. The UEs are equipped with $\NUE$ antennas and $\NUERF$ RF chains.
For simplicity, we consider that the antenna arrays at the BS and UEs are uniform linear arrays (ULAs) with half wavelength spacing. We will assume that the array response vector $\ba(\phi) \in \bbC^{N_{\rm a}}$ of a ULA with $N_{\rm a}$ elements at an incident angle $\phi$ has entries given by $\left[ \ba(\phi) \right]_{n_{\rm a}} = e^{-j\pi(n_{\rm a}-1)\sin(\phi)} $, for $n_{\rm a}=1,\ldots,N_{\rm a}$.

\subsection{Channel Models}
We leverage a sparse geometric channel model for the DL and UL channels operating at mmWave \cite{Heath2016JSTSP}. Specifically, the channel between the FD BS and the $i$-th DL UE at the $m$-th subcarrier $\HDL \in \bbC^{\NUE \times \NBST}$ can be written as
\begin{equation}\label{eq:Hdl}
    \HDL = \sum_{l=1}^{\LDL} \alphaDL e^{-j2\pi m\tauDL \deltaf} \aUE\big(\phiDL\big) \aBST^\rmH\big(\thetaDL\big),
\end{equation}
where $\LDL$ is the number of paths and $\deltaf$ is the subcarrier spacing. The complex channel gain, delay, angle-of-arrival (AoA) and angle-of-departure (AoD) of the $l$-th path of the $i$-th DL channel are denoted by $\alphaDL$, $\tauDL$, $\phiDL$ and $\thetaDL$, respectively. Furthermore, the array response vectors at UE arrays are represented by $\aUE(\phi) \in \bbC^{\NUE}$. Similarly, $\aBST(\phi) \in \bbC^{\NBST}$ denotes the array response at the TX array of the FD BS. Analogously, the channel between the $j$-th UL UE and FD BS at the $m$-th subcarrier $\HUL \in \bbC^{\NBSR \times \NUE}$ is defined as
\begin{equation}\label{eq:Hul}
    \HUL = \sum_{l=1}^{\LUL} \alphaUL e^{-j2\pi m\tauUL \deltaf} \aBSR\big(\phiUL\big) \aUE^\rmH\big(\thetaUL\big),
\end{equation}
where $\LUL$ is the number of paths. The complex channel gain, delay, AoA and AoD of the $l$-th path of the $j$-th UL channel are denoted by $\alphaUL$, $\tauUL$, $\phiUL$ and $\thetaUL$, respectively. The array response vector at the incident angle $\phi$ is denoted by $\aBSR(\phi) \in \bbC^{\NBSR}$ for the RX array of the FD BS.
We assume that the UE channels remain constant for the duration of the coherence time, which lasts several OFDM frames. Thus, we do not consider a doubly-selective channel model for the UEs, while we exploit that structure for the targets in order to estimate both the round-trip propagation delays and the Doppler shifts.
Since we consider a monostatic sensing scenario where the TX and RX arrays of the FD BS are placed very close to each other, the AoA and AoD of a target reflection are assumed to be the same \cite{Liyanaarachchi2021JCS,Barneto2022TCOM,Islam2022ICC,Islam2022GLOBECOM,Bayraktar2023CAMSAP,Bayraktar2024ICC,Talha2023Asilomar,Liu2023JSAC,He2023JSAC}. Hence, the target channel that includes $K$ targets at the $m$-th subcarrier of the $n$-th OFDM symbol $\Ht \in \bbC^{\NBSR \times \NBST}$ is expressed as
\begin{equation}\label{eq:Ht}
    \Ht = \sum_{k=0}^{K-1}\betat e^{j2\pi (n \Ts \fDk - m\taut \deltaf) } \aBSR(\thetat) \aBST^\rmH(\thetat),
\end{equation}
where $\Ts$ is the duration of an OFDM symbol such that $\Ts = 1/\deltaf + \Tcp$,  with $\Tcp$ the cyclic prefix duration. The complex radar gain, Doppler frequency, round-trip delay, and AoA/AoD of the $k$-th target are denoted by $\betat$, $\fDk$, $\taut$, and $\thetat$, respectively. The power of the radar gain is characterized with the radar-range equation as
\begin{equation}\label{eq:beta}
    |\betat|^2 = \frac{\lambda^2\varsigma_k}{(4\pi)^3 d_k^4},
\end{equation}
where $\lambda$ is the wavelength, while $\varsigma_k$ and $d_k$ are the radar cross-section and the range of the $k$-th target, respectively. Finally, the SI channel $\HSI \in \bbC^{\NBSR \times \NBST}$ is modeled with the near-field LoS model as \cite{Palacios2019GLOBECOM}
\begin{equation}\label{eq:HSI}
    \left[\HSI\right]_{p,q} = \frac{\gamma}{d_{pq}} e^{-j2\pi\frac{d_{pq}}{\lambda}},
\end{equation}
for $ p=1,\dots,\NBSR $ and $ q=1,\dots,\NBST $, where $d_{pq}$ is the distance between the $p$-th RX and $q$-th TX antenna, and $\gamma$ is the normalization constant to ensure $\norm{\HSI}_{\rmF}^2 = \NBST\NBSR$. Note that the SI channel can be modeled as a Rician channel with non-line-of-sight (NLoS) components caused by nearby scatterers. However, the Rician factor is usually very large as the TX and RX arrays are collocated, which makes the LoS component extremely dominant \cite{Satyanarayana2019TVT,Valcarce2020FD,Roberts2021TWC,Sheemar2022OJCOMS,Valcarce2022TWC}.

\subsection{Received Signals}
In this subsection, we introduce the models for the received signals at the DL UEs and FD BS. 
The TX side of the FD BS serves $\UDL$ DL UEs while also providing sensing capabilities. 
The number of data streams transmitted for each DL UE is denoted by $\NDLs$. 
Thus, the data stream transmitted to the $i$-th DL UE at the $m$-th subcarrier of the $n$-th symbol is represented by $\sDL \in \bbC^{\NDLs}$. Since the symbols and the streams of the UEs are independent, the data streams satisfy the following property $\Exp{\sDL\big[\sDLi\big]^\rmH} = \delta_{ii'}\frac{\PtDL}{\UDL\NDLs}\bI_{\NDLs}$, where $\PtDL$ is the DL transmit power. 
Additionally, the hybrid precoder at the FD BS for the $i$-th DL UE at the $m$-th subcarrier is expressed as $\FBS = \FBSRF\FBSBB \in \bbC^{\NBST\times\NDLs}$, with the frequency-flat and user-independent analog precoder $\FBSRF \in \bbC^{\NBST\times\NBSRF}$, and digital precoder $\FBSBB \in \bbC^{\NBSRF\times\NDLs}$. 
Furthermore, the hybrid combiner at the $i$-th DL UE for the $m$-th subcarrier is defined as $\WDL=\WDLRF\WDLBB \in \bbC^{\NUE\times\NDLs}$, with frequency-flat analog combiner $\WDLRF \in \bbC^{\NUE\times\NUERF}$ and digital combiner $\WDLBB \in \bbC^{\NUERF\times\NDLs}$. 
With these definitions, the received signal at the $i$-th DL UE for the $m$-th subcarrier of the $n$-th symbol $\yDL \in \bbC^{\NDLs}$ can be written as 
\begin{align}\label{eq:yDL}
    \yDL = \big[\WDL\big]^\rmH \HDL \FBS \sDL + \sum_{i'\neq i}^{\UDL} & \big[\WDL\big]^\rmH \HDL \FBSi \sDLi \nonumber \\ & + \big[\WDL\big]^\rmH \nDL,
\end{align}
where $\nDL \in \bbC^{\NUE}$ is the noise vector with covariance $\Exp{\nDL\big[\nDL\big]^\rmH} = \varDL\bI_{\NUE}$, being $\varDL$ the noise power. In \eqref{eq:yDL}, the MU interference is separated from the signal for the intended UE. We can simplify this expression as
\begin{equation}\label{eq:yDL2}
    \yDL = \big[\WDL\big]^\rmH \HDL \FBS \sDL + \nDLi,    
\end{equation}
where the interference-plus-noise term $\nDLi \in \bbC^{\NDLs}$ is defined as
\begin{equation}\label{eq:nDLi}
    \nDLi = \sum_{i'\neq i}^{\UDL} \big[\WDL\big]^\rmH \HDL \FBSi \sDLi \\ + \big[\WDL\big]^\rmH \nDL.  
\end{equation}

The received signal at the FD BS has a different form, since it includes the UL signals from $\UUL$ users, target reflections and SI. 
To provide simultaneous sensing and communication capabilities, and to represent the SI in the analog domain, we focus on the output of the analog combiner, which is denoted by $\WRF \in \bbC^{\NBSR\times\NBSRF}$. 
The data stream transmitted by the $j$-th UL UE is denoted by $\sUL\in\bbC^{\NULs}$. 
Like the DL data streams, UL data streams are independent for different users, which leads to the covariance $\Exp{\sUL\big[\sULj\big]^\rmH} = \delta_{jj'}\frac{\PtUL}{\NULs}\bI_{\NDLs}$, where $\PtUL$ is the transmit power of the $j$-th UL UE.
Finally, we define the hybrid precoder of the $j$-th UL UE as $\FUL=\FULRF\FULBB \in \bbC^{\NUE\times\NULs}$ with $\FULRF \in \bbC^{\NUE\times\NUERF}$  the frequency-flat analog precoder and $\FULBB \in \bbC^{\NUERF\times\NULs}$ the digital precoder. These definitions lead to the expressions of the received signals at the $m$-th subcarrier of the $n$-th symbol $\yBS \in \bbC^{\NULs}$, which can be written as
\begin{align}\label{eq:yBS}
    \yBS = & \sum_{j=1}^{\UUL} \WRF^\rmH \HUL \FUL \sUL + \sum_{i=1}^{\UDL} \WRF^\rmH \Ht \FBS \sDL \nonumber \\
    & \quad + \sqrt{\rho} \sum_{i=1}^{\UDL} \WRF^\rmH \HSI \FBS \sDL + \WRF^\rmH \nUL,
\end{align}
where $\rho$ is the SI channel power 
and $\nUL \in \bbC^{\NBSR}$ is the noise vector with covariance $\Exp{\nUL\nUL^\rmH} = \varUL\bI_{\NBSR}$, where $\varUL$ is the noise power at the BS. It is worthwhile to note that the received signal in \eqref{eq:yBS} serves for both the UL UEs and the radar processing. Moreover, the residual SI at this point should be suppressed to a level that does not cause LNAs to saturate. In the next section, we propose a combining architecture to separate UL UE signals, and also a radar processing framework to extract target reflections. To that end, we define the digital combiner for the $j$-th UL UE at the $m$-th subcarrier as $\WBB \in \bbC^{\NBSRF\times\NULs}$,  and the related processed signal $\yUL \in \bbC^{\NULs}$ as
\begin{equation}\label{eq:yUL}
    \yUL = \big[\WBB\big]^\rmH\WRF^\rmH \HUL \FUL \sUL + \nULj,
\end{equation}
where the interference-plus-noise term $\nULj \in \bbC^{\NULs}$ is defined as
\begin{multline}\label{eq:nULj}
    \nULj = \big[\WBB\big]^\rmH \Bigg( \sum_{j'\neq j}^{\UUL} \WRF^\rmH \HULj \FULj \sULj \\ 
    + \sum_{i=1}^{\UDL} \WRF^\rmH \Ht \FBS \sDL
    \\ + \sqrt{\rho} \sum_{i=1}^{\UDL} \WRF^\rmH \HSI \FBS \sDL + \WRF^\rmH \nUL \Bigg).
\end{multline}

\section{Precoder and Combiner Design}
In this section, we propose a hybrid precoding and combining framework to support simultaneous monostatic sensing and FD MU communication for mmWave MIMO systems. To that end, we first define the communication and sensing metrics that we will consider, which are the sum spectral efficiency and the radar gain at the target angle. Then, we define the optimization problem to be solved to design hybrid precoders and combiners that provide high sum spectral efficiency for all the UEs while maintaining high radar gain at the target angle under the SI and analog hardware constraints. Using the DL received signal expression in \eqref{eq:yDL2}, the spectral efficiency of the $i$-th DL at the $m$-th subcarrier is expressed as
\begin{align}
    \rateDL = \log_{2} \Big| & \bI_{\NDLs} + \tfrac{\PtDL}{\UDL\NDLs}\Big(\RnDLi\Big)^{-1} \big[\WDL\big]^\rmH \nonumber\\ 
    & \:\: \times \HDL \FBS \big[\FBS\big]^\rmH \big[\HDL\big]^\rmH \WDL  \Big|,
\end{align}
where the covariance matrix of the interference-plus-noise vector is defined as $\RnDLi = \Exp{\nDLi [\nDLi]^\rmH} \in \bbC^{\NDLs\times\NDLs}$. Analogously, using the received signal expression in \eqref{eq:yUL}, the spectral efficiency of the $j$-th UL UE at the $m$-th subcarrier can be written as
\begin{align}
    \rateUL = & \log_{2} \Big| \bI_{\NULs} + \tfrac{\PtUL}{\NULs} \Big(\RnULj\Big)^{-1} \big[\WBB\big]^\rmH \WRF^\rmH \nonumber \\
    & \quad \times \HUL \FUL \big[\FUL\big]^\rmH \big[\HUL\big]^\rmH \WBB\WRF  \Big|,
\end{align}
where the covariance matrix of the interference-plus-noise vector is defined as $\RnULj = \Exp{\nULj [\nULj]^\rmH} \in \bbC^{\NULs\times\NULs}$. 

For the sensing metric, we utilize the radar gain achieved with the precoders and the analog combiner used at the FD BS. As explained previously, we focus on tracking a single target at each OFDM frame. The subscript of the tracked target is conveniently denoted by $k=0$ in the target channel expression \eqref{eq:Ht}. Hence, the TX radar gain at the target angle $\theta_0$ with the precoder $\FBS$ is defined as
\begin{equation}
	\GT(\theta_0) = \left| \aBST^\rmH(\theta_0) \big[\FBS\big]_{:,\ns} \right|,
\end{equation}
for $\ns=1,\dots,\NDLs$. Similarly, the RX radar gain at the target angle $\theta_0$ can be defined as 
\begin{equation}
    \GR(\theta_0) = \Big|[\WRF]_{:,\nRF}^\rmH \aBSR(\theta_0)\Big|,
\end{equation}
for $\nRF=1,\dots,\NBSRF$.

Aside from the communication and sensing metrics, we also need to define the SI and analog constraints. First, the residual SI at the output of the analog combiner of the FD BS, which is given in \eqref{eq:yBS}, should be mitigated. This constraint can be written as
\begin{equation}\label{eq:SI}
    \left[\WRF\right]^\rmH  \HSI \FBS = \boldsymbol{0},
\end{equation}
for $m=0,\ldots,M-1$ and $i=1,\ldots,\UDL$. The analog constraint is related to the analog precoders/combiners that are realized with networks of phase shifters. That is, the entries of the analog precoders/combiners should have unit modulus entries. At this stage, we formulate the overall problem while ignoring hybrid analog/digital structure for the precoders at the FD BS, combiners at the DL UEs and precoders at the UL UEs. We enforce unit modulus entries only on the analog combiners at the FD BS since the SI mitigation should be realized at this stage. The hybrid analog/digital architecture will be enforced on the remaining precoders/combiners later. Considering the mentioned definitions, we write the overall optimization problem as the maximization of the sum spectral efficiency of all UEs while constraining the radar gains to be above predefined thresholds as
\begin{equation}\label{eq:problem_overall}
    \begin{aligned}
    & \underset{\mathclap{\substack{\big\{\FBS, \WDL\big\}_{m,i}, \WRF,\\ \big\{\FUL, \WBB\big\}_{m,j}}}}{\text{maximize}} \quad\quad & & \sum_{i=1}^{\UDL}\sum_{m=0}^{M-1} \rateDL + \sum_{j=1}^{\UUL}\sum_{m=0}^{M-1} \rateUL\\
    & \text{subject to} & &  \GT(\theta_0) \geq \tauT, \: \forall m, \ns, i\\
    & & & \GR(\theta_0) \geq \tauR, \: \forall \nRF, \\
    & & & \left[\WRF\right]^\rmH  \HSI \FBS = \boldsymbol{0}, \: \forall m,i \\
    & & & \left|\left[\WRF\right]_{p,q}\right| = 1, \: \forall p,q, \\    
    & & & \big\|\FBS\big\|_\rmF^{2} = \NDLs, \: \forall m,i, \\
    & & & \big\|\FUL\big\|_\rmF^{2} = \NULs, \: \forall m,j,
\end{aligned}
\end{equation}
where $\tauT$ and $\tauR$ are the TX and RX gain thresholds, respectively, whereas the last constraint ensures the power normalization for the precoders. This problem, even when the hybrid architecture is not enforced, is highly non-linear due to the coupling of the variables in the objective function. Furthermore, the unit modulus constraint for the analog combiner at the BS makes the problem non-convex. To be able to solve this problem, we employ an alternating optimization method to create subproblems which consider a single variable to be optimized while the others are fixed. Furthermore, we utilize relaxation techniques to obtain convex problems at each stage. In the rest of the section, we introduce our proposed design for each precoder/combiner.

\subsection{Hybrid Precoder Design at the FD BS}
We start designing the precoders at the FD BS by firstly omitting the hybrid analog/digital architecture. Considering that the other variables are fixed, the design problem can be written as
\begin{equation}\label{eq:problem_FBS}
    \begin{aligned}
    & \underset{\big\{\FBS\big\}_{m,i}}{\text{maximize}} & & \sum_{i=1}^{\UDL}\sum_{m=0}^{M-1} \rateDL \\
    & \text{subject to} & &  \GT(\theta_0) \geq \tauT, \: \forall m, \ns, i\\
    & & & \left[\WRF\right]^\rmH  \HSI \FBS = \boldsymbol{0}, \: \forall m,i \\
    & & & \big\|\FBS\big\|_\rmF^{2} = \NDLs, \: \forall m,i.
\end{aligned}
\end{equation}
This problem can be solved in parallel for all the subcarriers since we consider a fully digital architecture. Moreover, the power constraint can be omitted and enforced later. We further separate the optimization of the communication and sensing metrics. In other words, we design a precoder for communication $\FBScom \in \bbC^{\NBST \times \NDLs}$ and another precoder for sensing $\FBSrad \in \bbC^{\NBST \times \NDLs}$. Then, we coherently combine these precoders to obtain the final precoder as
\begin{equation}\label{eq:FBS}
    \FBS = \kappa_1\FBScom + (1-\kappa_1)\FBSrad,
\end{equation}
where $\kappa_1$ is the parameter that controls the trade-off between communication and sensing performance \cite{Elbir2023SPM}. Note that the communication and sensing precoders satisfy the power constraint. Additionally, we normalize the final precoder such that $\big\|\FBS\big\|_\rmF^{2} = \NDLs$. Furthermore, the selection of $\kappa_1$ depends on the TX gain constraint. We utilize a simple search approach to find the largest $\kappa_1$ value that satisfies the TX gain constraint for all the streams of the power normalized final precoder.

An important observation related to \eqref{eq:FBS} is that, since both the communication and radar precoders consider the SI constraint, the coherent combination also has SI suppression capability. In the rest of the subsection, we describe the design of SI-aware communication and sensing precoders, as well as the hybrid factorization considered for the final precoders.

\subsubsection{Communication Precoder Design}
The communication precoder design problem for the $m$-th subcarrier with the SI mitigation constraint is expressed as
\begin{equation}\label{eq:problem_FBScom}
    \begin{aligned}
    & \underset{\big\{\FBScom\big\}_{i}}{\text{maximize}} & & \sum_{i=1}^{\UDL} \rateDL \\
    & \text{subject to} & & \left[\WRF\right]^\rmH  \HSI \FBScom = \boldsymbol{0}, \: \forall i, \\
    & & & \big\|\FBScom\big\|_\rmF^{2} = \NDLs, \: \forall i.
\end{aligned}
\end{equation}
It is very challenging to optimize the cost function in \eqref{eq:problem_FBScom}, even without the SI mitigation constraint. The main reason is the MU interference observed in the expression \eqref{eq:nDLi}. Ideally, we would also like to completely mitigate the MU interference terms, which would be similar to the SI mitigation constraint. Solving this problem is still challenging since the precoders of all DL UEs are coupled. In \cite{Sadek2007TWC}, the authors propose SLNR-based precoding to solve the MU precoding problem. We will adopt a similar approach, however, we will include SI in the leakage term. 

\begin{figure*}[!t]
\normalsize
\begin{equation}\label{eq:SLNR}
    \SLNR = \frac{
    \Exp{ \norm{\big[\WDL\big]^\rmH \HDL \FBScom \sDL}^2 }
    }{
    \sum_{i'\neq i}\Exp{\norm{\big[\WDLi\big]^\rmH \HDLi \FBScom \sDL}^2 } + \Exp{\norm{\sqrt{\rho} \WRF^\rmH \HSI \FBScom \sDL }^2} + \Exp{\norm{\big[\WDL\big]^{\rmH}\nDLi}^2}
    }
\end{equation}
\hrulefill
\vspace{-10pt}
\end{figure*}

The TX signal of the $i$-th DL UE causes leakage for the received signal of the $i'$-th DL UE in the form of $\big[\WDLi\big]^\rmH \HDLi \FBScom \sDL$ as can be deduced from \eqref{eq:nDLi}. Similarly, due to the SI channel, the TX signal of the $i$-th DL UE causes leakage for the received signal at the FD BS in the form of $\sqrt{\rho} \WRF^\rmH \HSI \FBScom \sDL$, which can be seen from \eqref{eq:yBS}. Considering the noise in the received signal, the SLNR at the $m$-th subcarrier for the $i$-th DL UE is given in \eqref{eq:SLNR} on the top of next page. We simplify the given expression by using the trace operator and the fact that the data streams are independently distributed. The numerator in \eqref{eq:SLNR} can be rewritten as
\begin{align}
    & \Exp{\norm{\big[\WDL\big]^\rmH \HDL \FBScom \sDL}^2 } \nonumber\\
    &\quad\quad = \Exp{\big[\sDL\big]^{\rmH} \big[\FBScom\big]^{\rmH} \Bcov \FBScom \sDL } \nonumber\\
    &\quad\quad = \trace{\Exp{\big[\sDL\big]^{\rmH} \big[\FBScom\big]^{\rmH} \Bcov \FBScom \sDL }} \nonumber\\
    &\quad\quad = \trace{\Exp{\big[\sDL\big]^{\rmH} \sDL} \big[\FBScom\big]^{\rmH} \Bcov \FBScom} \nonumber\\
    &\quad\quad = \frac{\PtDL}{\UDL\NDLs} \trace{\big[\FBScom\big]^{\rmH} \Bcov \FBScom},
\end{align}
where $\Bcov = \big[\HDL\big]^{\rmH} \WDL \big[\WDL\big]^\rmH \HDL \in \bbC^{\NBST \times \NBST}$. We can simplify the MU and SI leakage terms by following the same steps as
\begin{multline}
    \Exp{\norm{\big[\WDLi\big]^\rmH \HDLi \FBScom \sDL}^2 } \\ = \frac{\PtDL}{\UDL\NDLs} \trace{\big[\FBScom\big]^{\rmH} \Bcovi \FBScom},
\end{multline}
\begin{multline}
    \Exp{\norm{\sqrt{\rho} \WRF^\rmH \HSI \FBScom \sDL }^2} \\ = \frac{\PtDL}{\UDL\NDLs} \trace{\big[\FBScom\big]^{\rmH} \bC \FBScom},
\end{multline}
where $\bC = \rho \HSI^\rmH \WRF \WRF^\rmH \HSI \in \bbC^{\NBST \times \NBST} $. Finally, the noise term can be simplified as
\begin{align}
    \Exp{\norm{\big[\WDL\big]^{\rmH}\nDLi}^2} & = \Exp{\big[\nDLi\big]^{\rmH} \WDL \big[\WDL\big]^\rmH \nDLi} \nonumber\\
    &\hspace{-10pt} = \trace{\big[\WDL\big]^\rmH \Exp{\nDLi \big[\nDLi\big]^{\rmH}} \WDL} \nonumber\\
    &\hspace{-10pt} = \varDL \trace{\big[\WDL\big]^\rmH \WDL} \nonumber\\
    &\hspace{-10pt} = \NDLs \varDL,
\end{align}
where we assumed $\trace{\big[\WDL\big]^\rmH \WDL} = \NDLs$ without loss of generality. Utilizing the simplifications given above, the SLNR expression in \eqref{eq:SLNR} can be rewritten as
\begin{align}\label{eq:SLNR2}
    &\SLNR \nonumber\\ &= \frac{
    \trace{\big[\FBScom\big]^{\rmH} \bBcov \FBScom}}{
    \trace{\big[\FBScom\big]^{\rmH} \Big(\sum_{i'\neq i}\bBcovi \!+ \! \bar{\bC} \Big) \FBScom} + \NDLs \varDL} \nonumber\\
    & = \! \frac{
    \trace{\big[\FBScom\big]^{\rmH} \bBcov \FBScom}}{
    \trace{\big[\FBScom\big]^{\rmH} \Big(\sum_{i'\neq i}\bBcovi \!+\! \bar{\bC} + \varDL \bI_{\NBST} \Big) \FBScom}},
\end{align}
where we defined $\bBcov=\frac{\PtDL}{\UDL\NDLs}\Bcov$ and $\bar{\bC}=\frac{\PtDL}{\UDL\NDLs}\bC$. Instead of solving the problem in \eqref{eq:problem_FBScom}, we propose to maximize SLNR for each DL UE. It is important to note that while the problem in $\eqref{eq:problem_FBScom}$ aims to maximize the spectral efficiency of all DL UEs simultaneously, maximization of SLNR is considered separately. The maximization of the SLNR of the $i$-th DL UE at the $m$-th subcarrier is formulated as
\begin{equation}\label{eq:problem_FBScom_SLNR}
    \begin{aligned}
    & \underset{\FBScom}{\text{maximize}} & & \SLNR \\
    & \text{subject to} & & \big\|\FBScom\big\|_\rmF^{2} = \NDLs.
\end{aligned}
\end{equation}
This is a well-known generalized eigenvalue problem which is studied extensively in \cite{Ghojogh2019arXiv}. This problem is equivalently expressed as
\begin{equation}
    \bBcov \FBScom = \Bigg(\! \sum_{i'\neq i}\bBcovi + \bar{\bC} + \varDL \bI_{\NBST} \!\! \Bigg) \FBScom \boldsymbol{\Lambda}_{m}^{(i)},
\end{equation}
where the columns of $\FBScom$ are the generalized eigenvectors, $\boldsymbol{\Lambda}_{m}^{(i)} \in \bbC^{\NDLs\times\NDLs}$ is a diagonal matrix that contains the corresponding eigenvalues on its diagonal entries. Note that this solution diagonalizes the term inside the trace in the denominator of the SLNR expression in \eqref{eq:SLNR2}, which leads to the mitigation of inter-stream interference. Furthermore, the SLNR expression only requires the combined channel, i.e., $\big[\WDL\big]^\rmH\HDL$, instead of the full dimensional DL channels. This is especially useful if the channels are estimated at the DL UEs since the feedback overhead is reduced with using the combined channels. Lastly, $\FBScom$ satisfies the power constraint since the eigenvectors have unit norm.

\subsubsection{Sensing Precoder Design}
We design the sensing precoder by using the same methodology as the communication precoder. Since the TX radar gain constraint is satisfied with the coherent combination in \eqref{eq:FBS}, our goal is to find a precoder that has high TX radar gain while resulting in low residual SI. Since we utilize the common streams for joint communication and sensing, the sensing precoders should also consider the MU interference. Hence, we use the same SLNR expression in \eqref{eq:SLNR2} by only changing the numerator which contains the intended signal power.

In this paper, we are interested in tracking a single target, $k=0$, per frame. Therefore, we can take the power of the TX radar gain as the maximization term
\begin{equation}
    \big(\GT\!(\theta_0)\big)^2 \!\!=\! \big[\FBSrad\big]_{:,\ns}^\rmH \! \aBST(\theta_0) \aBST^\rmH(\theta_0) \big[\FBSrad\big]_{:,\ns}\!\!.
\end{equation}
The rank of the inner matrix $\aBST(\theta_0) \aBST^\rmH(\theta_0)$ is 1, thus, it is not possible to find different eigenvectors for different streams. Consequently, we optimize a single vector $\fBSrad \in \bbC^{\NBST}$, and repeat it for the columns of $\FBSrad$. This approach does not enhance the inter-stream interference as long as one of the dominant paths of the UE does not overlap with $\theta_0$. Even in that case, the inter-stream interference increase is limited as small $\kappa_1$ values are sufficient to achieve the required TX radar gain due to overlapping. With the given definition, the SLNR for the sensing beam is given as
\begin{align}\label{eq:SLNRrad}
    \SLNRrad = \frac{
    \big[\fBSrad\big]^{\rmH} \aBST(\theta_0) \aBST^\rmH(\theta_0) \fBSrad }{
    \big[\fBSrad\big]^{\rmH} \Big(\sum_{i'\neq i}\bBcovi \!+\! \bar{\bC} + \varDL \bI_{\NBST} \Big) \fBSrad}.
\end{align}
The maximization of the sensing SLNR can be formulated by following the same approach in \eqref{eq:problem_FBScom_SLNR}, and the generalized eigenvalue solution can be used. It is important to note that it is possible to consider the received signals from other targets ($k>0$) as leakage. We do not utilize this approach since the target returns are not strong and the beams are narrow. Furthermore, this approach would require the knowledge of radar channel coefficients $\betat$ which may not be available, especially in a simultaneous UL communication framework. If clutter is considered, this approach may be meaningful due to the availability of reliable clutter estimates. Although we do not consider it in this work, our approach can be easily adapted to scenarios with clutter.  

\subsubsection{Hybrid Factorization}
The solution provided in \eqref{eq:FBS} is suitable for a fully digital architecture. Since mmWave systems require a hybrid analog/digital architecture to reduce power consumption, we utilize a hybrid factorization algorithm to find the analog and digital precoders from the fully digital ones. The challenging part of this problem is that the analog precoder $\FBSRF$ should be common for all DL UEs and subcarriers. The hybrid factorization problem for MU mmWave systems is formulated as \cite{Coma2018JSTSP,Fernandez2018SPAWC}
\begin{equation}\label{eq:hybrid}
	\begin{aligned}
		& \underset{\mathclap{\FBSRF,\big\{\FBSBB\big\}_{m,i}}}{\text{minimize}} \quad & & \sum_{i=1}^{\UDL} \sum_{m=0}^{M-1} \norm{\FBS-\FBSRF\FBSBB}_\rmF^{2}\\
		& \text{subject to} & & \left| \left[\FBSRF\right]_{p,q} \right| = 1, \: \forall p,q,\\
        & & & \norm{\FBSRF\FBSBB}_\rmF^{2} = \NDLs, \: \forall m, i.
	\end{aligned}
\end{equation}
We exploit the low-complexity greedy hybrid precoding solution in \cite{Fernandez2018SPAWC}. This is an iterative solution such that the digital precoders are updated with the least-squares solution while a greedy approach is adopted for the update of the analog precoder. This solution does not preserve the leakage suppression, which is especially critical for the FD system. Thus, we further suppress the SI with the analog combiner at the FD BS.

\subsection{Hybrid Combiner Design at DL UEs and Hybrid Precoder Design at UL UEs}
The UEs have the knowledge of their own channels while they are agnostic the channels of other UEs. Thus, MU interference suppression is not considered at UEs. In this case, it is reasonable to utilize the optimum single-user combiners/precoders. For the combiner of the $i$-th DL UE at the $m$-th subcarrier, we use the first $\NDLs$ left singular vectors of the channel $\HDL$. Then, we use a hybrid factorization technique to find the frequency-flat analog combiner and digital combiners. The hybrid factorization problem is in the same form as \eqref{eq:hybrid} with only one user. Hence, we can use the same algorithm from \cite{Fernandez2018SPAWC}. Similarly we design the precoder of the $j$-th UL UE at the $m$-th subcarrier by taking the $\NULs$ left singular vectors of the channel $\HUL$, which is followed by the same hybrid factorization algorithm.



\subsection{Analog Combiner Design at FD BS}
The combiner at the FD BS should simultaneously provide communication for the UL UEs and high RX radar gain. The SI mitigation requirement at the output of the analog combiner makes the problem uniquely challenging, mainly due to the non-convexity of the unit modulus constraint on the entries. In this subsection, we reformulate the problem and utilize convex relaxation techniques to design the analog combiner.

We firstly note that the SI mitigation is the primary goal at this stage since the hybrid factorization applied to the precoders at the FD BS reduces the SI suppression capability. Furthermore, while the analog combiner should jointly provide communication and sensing services, the digital stage will be utilized for suppressing the MU detection and radar processing. On one hand, the RX radar gain constraint defined in the initial optimization problem is utilized for sensing purposes. On the other hand, we propose to employ a similar approach for communication. To that end, we need to find a set of vectors that are useful for communication. This can be achieved by using the eigenvectors that correspond to the $\NULs$ largest eigenvalues of the precoded channel covariance matrix $\frac{1}{M}\sum_{m}\HUL\FUL\big[\FUL\big]^\rmH\big[\HUL\big]^\rmH$ of each UL UE as done in \cite{Zhang2019IET}. Note that the precoded channels, i.e., $\HUL\FUL$, are available at the FD BS instead of the full dimensional channels, reducing the channel estimation and feedback overhead. We denote the matrix that consists of the eigenvectors of the channels by $\WRFcom \in \bbC^{\NBSR \times \NBSRF}$. If $\NBSRF > \UUL\NULs$, the remaining columns can be constructed as a random linear combination of the first $\UUL\NULs$ columns. We normalize the norm of each column to $\sqrt{\NBSR}$ to match the norm of the columns of an analog combiner. With this definition, similar to the RX radar gain, we define a communication gain as
\begin{equation}
    \Gcom = \Big|[\WRF]_{:,\nRF}^\rmH [\WRFcom]_{:,\nRF}\Big|,
\end{equation}
for $\nRF=1,\dots,\NBSRF$. We formulate the optimization problem for the analog combiner $\WRF$ as the minimization of the total SI with RX radar gain and communication gain constraints. Given the precoders at the FD BS and at the UL UEs, and considering the unit modulus constraint, this problem can be expressed as
\begin{equation}\label{eq:combiner_problem}
    \begin{aligned}
    & \underset{\WRF}{\text{minimize}} & & \sum_{i=1}^{\UDL} \sum_{m=0}^{M-1} \norm{\WRF^\rmH \HSI \FBS}_{\rmF}^{2}\\
    & \text{subject to} & & \GR(\theta_{0}) \geq \tauR, \: \forall \nRF, \\
    & & & \Gcom \geq \taucom, \: \forall \nRF, \\
    & & & \left|\left[\WRF\right]_{p,q}\right| = 1, \: \forall p,q,
\end{aligned}
\end{equation}
where $\taucom$ is the communication gain threshold. This problem is non-convex due to the unit modulus constraint. Additionally, the gain constraints are non-convex, which can be circumvented by changing the gain functions to convex gain loss functions that are defined as
\begin{equation}
    \GRloss(\theta_0) = \Big|\NBSR - [\WRF]_{:,\nRF}^\rmH \aBSR(\theta_0)\Big|,
\end{equation}
\begin{equation}
    \Gcomloss = \Big|\NBSR - [\WRF]_{:,\nRF}^\rmH [\WRFcom]_{:,\nRF}\Big|.
\end{equation}
In the modified optimization problem, we limit the gain loss functions to be below thresholds instead of forcing the gains to be above a threshold. The thresholds for the RX gain loss and the communication gain loss are expressed as $\NBSR - \tauR$ and $\NBSR - \taucom$, respectively. These constraints are not exactly equal to the initial gain constraint, however, it can be shown that they provide an upper bound. Hence, it is meaningful to use the gain loss constraints. 

To overcome the non-convexity of the unit modulus constraint, we limit the deviation of the entries of the combiner $\WRF$ from the unit circle as done in \cite{Bayraktar2023CAMSAP,Bayraktar2024ICC}. We utilize a random block coordinate descent method to solve the convex problem in an iterative manner \cite{Nesterov2012SIAM}. At each iteration, a certain number of randomly selected entries are optimized. The iterations are terminated when the cost function converges. Additionally, the overall evolution of the combiner is limited to prevent missing the local minima. This approach is analogous to the step size in gradient descent algorithms. The convex problem solved at the $t$-th iteration is expressed as
\begin{equation}\label{eq:problem2}
    \begin{aligned}
    & \underset{\left[\WRF\right]_{\cI_t}}{\text{minimize}} & & \sum_{i=1}^{\UDL} \sum_{m=0}^{M-1} \norm{\WRF^\rmH \HSI \FBS}_{\rmF}^{2}\\
    & \text{subject to} & & \GRloss(\theta_0) \leq (\NBSR - \tauR),\forall \nRF\\
    & & & \Gcomloss \leq (\NBSR - \taucom),\forall \nRF\\
    & & & \left|\left[\WRF\right]_{p,q}\right| - 1 \leq \epsilon_1, \: (p,q) \in \cI_t, \\
    & & & \norm{\left[\WRF\right]_{\cI_t} - \left[\WRF^{(t-1)}\right]_{\cI_t}  } \leq \epsilon_2,
\end{aligned}
\end{equation}
where $\epsilon_1$ and $\epsilon_2$ are the threshold values. Moreover, the set of randomly selected indices of the analog combiner at the $t$-th iteration is denoted by $\cI_t$. The entries to be optimized at the $t$-th iteration are represented by $ \left[\WRF\right]_{\cI_t} \in \bbC^{|\cI_t|} $, whereas the analog combiner optimized in the $(t-1)$-th iteration is denoted by $\WRF^{(t-1)}$. At the end of each iteration, the entries are normalized to unity to satisfy the unit modulus constraint. The problem \eqref{eq:problem2} is solved with the convex solver CVX \cite{Grant2014cvx}.

The provided iterative solution requires an initial combiner. To that end, we resort to the coherent combination of communication and sensing combiners as
\begin{equation}
    \WRF^{\rm init} = \kappa_2 \WRFcom + (1-\kappa_2) \WRFrad,
\end{equation}
where the sensing combiner $\WRFrad \in \bbC^{\NBSR \times \NBSRF}$ is constructed as $\WRFrad = \big[\aBSR(\theta_0), \ldots, \aBSR(\theta_0)\big]$. Furthermore, $\kappa_2$ is the trade-off parameter for communication and sensing. We normalize the entries of the initial combiner to unity to satisfy the unit modulus constraint.  

\subsection{Digital Combiner Design at FD BS}
The digital combiner design for the UL UEs is more straightforward compared to the other precoders and combiners. At this stage, we assume that SI is jointly suppressed by the precoders and the analog combiner utilized at the FD BS. Furthermore, we assume that the MU interference dominates the other interference sources, i.e., target reflections and the residual SI in \eqref{eq:nULj}. Assuming that the precoders employed at UL UEs are known, we utilize the linear minimum mean square error (LMMSE) combiner. With the given assumptions and the expression in \eqref{eq:yUL}, the LMMSE combiner of the $j$-th UL UE at the $m$-th subcarrier is derived as
\begin{align}\label{eq:WBB}
    \WBB = \Bigg(\sum_{j'=1}^{\UUL} \frac{\PtUL}{\NULs} \WRF^\rmH \HULj \FULj \big[\FULj\big]^\rmH \nonumber\\ \times \big[\HULj\big]^\rmH \WRF + \varUL \WRF^\rmH \WRF \Bigg)^{-1} \WRF^\rmH \HUL \FUL,
\end{align}
where we utilized the independence of the data streams of the UL UEs. The SI and target interference are not included in this expression, as previously explained. If the interference from these sources is strong, there are alternative solutions such as digital interference cancellation for SI. Since the DL transmit symbols are known by the FD BS, the interference can be efficiently cancelled by subtracting the interference terms from the received signal in \eqref{eq:yUL}. The estimates of the target parameters can be utilized to construct the target channel for this purpose.

\subsection{Digital Combiner Design for Radar Processing}
In this subsection, we design a digital combiner for radar processing. Since we are only interested in a single target, we can rewrite the received signal in \eqref{eq:yBS} by separating the signal of the intended target as
\begin{align}\label{eq:yBSrad}
    \yBS \!=\! \bar{\beta}_{0,m,n} \WRF^\rmH \aBSR(\theta_0) \! \sum_{i=1}^{\UDL} \aBST^\rmH(\theta_0) \FBS \sDL \! + \! \nULo,
\end{align}
where $\bar{\beta}_{0,m,n} = \beta_0 e^{j2\pi (n \Ts f_{{\rm D}, 0} - m\tau_0 \deltaf) }$, and $\nULo \in \bbC^{\NBSRF}$ contains all the terms in the received signal except for the signal related to the intended target. We assume that the interference from the other targets is much weaker than the signal of the intended target. The main reason is that we design the analog combiner $\WRF$ to have a high gain at the angle $\theta_0$. Since we have narrow beams at mmWave, the interference from other sources will be limited. We also assume that the SI is suppressed at this stage, similar to what we assumed for the digital combiners of the UL UEs. With the given assumptions, we can use the same radar combiner designed for the $m$-th subcarrier for all symbols as the dependency on the symbol index $n$ is removed by neglecting the returns of unintended targets. We define the radar combiner for the $m$-th subcarrier as $\bw_{m} \in \bbC^{\NBSRF}$. This combiner is designed by using the MVDR beamformer approach \cite{VanTrees2002}. The cost function of the problem is the minimization of the residual noise term expressed as
\begin{align}
    \Exp{\left|\bw_{m}^\rmH \bnULo \right|^2} & = \bw_{m}^\rmH \Exp{\bnULo [\bnULo]^\rmH} \bw_{m} \nonumber \\
    & = \bw_{m}^\rmH \bR_{\bnULom} \bw_{m},
\end{align}
where $\bnULo \in \bbC^{\NBSRF}$ is the interference vector without the target interference and SI, and $\bR_{\bnULom} \in \bbC^{\NBSRF \times \NBSRF}$ is its covariance matrix, which can be written as
\begin{align}
    \bR_{\bnULom} = \sum_{j=1}^{\UUL} \frac{\PtUL}{\NULs} \WRF^\rmH \HUL \FUL \big[\FUL\big]^\rmH \nonumber\\ \times \big[\HUL\big]^\rmH \WRF + \varUL \WRF^\rmH \WRF.
\end{align}
The MVDR beamformer problem can be formulated as
\begin{equation}\label{eq:MVDR}
    \begin{aligned}
    & \underset{\bw_{m}}{\text{minimize}} & & \bw_{m}^\rmH \bR_{\bnULom} \bw_{m}\\
    & \text{subject to} & & \bw_{m}^\rmH \WRF^\rmH \aBSR(\theta_0) = 1.
\end{aligned}
\end{equation}
The optimum solution to this problem is given as \cite{VanTrees2002}
\begin{equation}
    \bw_{m} = \frac{\bR_{\bnULom}^{-1} \WRF^\rmH \aBSR(\theta_0)}{\aBSR^\rmH(\theta_0) \WRF \bR_{\bnULom}^{-1} \WRF^\rmH \aBSR(\theta_0)}.
\end{equation}
One important observation related to the proposed architecture is that the inverse of the covariance matrix $\bR_{\bnULom}$ used in the MVDR beamformer is exactly the same as the inverted matrix in \eqref{eq:WBB}, which is the result of ignoring the residual SI and target interference. Therefore, the same matrix inverse is calculated for communication and radar processing at the RX side of the FD BS leading to reduced complexity. Furthermore, it is important to note that the MVDR beamformer suppresses the UL interference enabling sensing in the presence of UL communication.


\section{Parameter Estimation for Monostatic Sensing}
In this section, we introduce the techniques for target parameter estimation. Since we consider the tracking stage, the erroneous initial estimate of the target angle is assumed to be available. The precoders and the analog combiner at the FD BS are designed by using the initial angle estimate. We first introduce the angle refinement procedure approach based on well-known MUSIC algorithm. Then, we explain the range-velocity estimation stage which is based on OFDM radar processing.

\subsection{Angle Estimation}
We have a strong target reflection as the precoders and the analog combiner at FD BS are designed with the knowledge of the initial angle estimate. Although simultaneous reception of the UL signals and target reflections complicates the angle detection, knowledge of the uplink channels can be leveraged for the subspace-based angle estimation methods, such as MUSIC. We first calculate the covariance matrix of the received signal at the output of the analog combiner $\WRF$ in \eqref{eq:yBSrad}. Similar to what we assumed in the previous section, we neglect the residual SI and target interference. In this case the covariance matrix of the received signal is assumed to have UL signals and correlated noise terms. Thus, the covariance matrix $\bR_{\by} \in \bbC^{\NBSRF \times \NBSRF}$ is approximated as
\begin{align}\label{eq:Ry}
    & \bR_{\by} = \Exp{\yBS\yBS^\rmH} \nonumber \\
    & \approx |\tilde{\beta}_0|^2 \WRF^\rmH \aBSR(\theta_0)\aBSR^\rmH(\theta_0) \WRF \nonumber \\
    & \quad + \sum_{j=1}^{\UUL} \frac{\PtUL}{\NULs} \WRF^\rmH \Exp{ \HUL \FUL \big[\FUL\big]^\rmH \big[\HUL\big]^\rmH} \WRF \nonumber \\
    & \quad \quad + \varUL \WRF^\rmH \WRF,
\end{align}
where the effective gain of the target return is expressed as
\begin{equation}
    |\tilde{\beta}_0|^2 =\frac{\PtDL|\beta_0|^2}{\UDL\NDLs}\aBST^\rmH(\theta_0) \Exp{\FBS \big[\FBS\big]^\rmH} \aBST(\theta_0).
\end{equation}
We calculate the sample covariance matrix based on our observations as $\bR_{\by} = \frac{1}{MN} \sum_{m=0}^{M-1} \sum_{m=0}^{N-1} \yBS\yBS^\rmH $. Similarly, we can calculate the sample covariance matrix of the UL signals since we assume to have the knowledge of the precoded UL channels at the BS. Therefore, we can calculate the UL signal-free sample covariance matrix $\bar{\bR}_{\by} \in \bbC^{\NBSRF\times\NBSRF}$ as
\begin{align}
    &\bar{\bR}_{\by} = \frac{1}{MN} \sum_{m=0}^{M-1} \sum_{n=0}^{N-1} \Bigg( \yBS\yBS^\rmH \nonumber \\
    & - \sum_{j=1}^{\UUL} \frac{\PtUL}{\NULs} \WRF^\rmH \HUL \FUL \big[\FUL\big]^\rmH \big[\HUL\big]^\rmH \WRF \Bigg).
\end{align}
This expression contains the covariance matrix of the intended target and the correlated noise. We can estimate the most dominant angle by using the MUSIC algorithm, however, we need to have white noise to operate with this algorithm. Thus, we use the Cholesky decomposition of the noise covariance matrix $\bL\bL^\rmH = \WRF^\rmH \WRF$ and transform the UL signal-free covariance matrix to $\bL^{-1} \bar{\bR}_{\by} \big[\bL^{-1}\big]^\rmH$, which whitens the noise. Then, we can use the MUSIC algorithm in the beamspace to find the peak in the angular pseudospectrum \cite{VanTrees2002}. The advantage of the tracking is that we can limit the angular search range to the vicinity of the initial angle estimate. The refined angle estimate of the target is denoted by $\hat{\theta}_0$.

\subsection{Range-Velocity Estimation}
In the context of ISAC, OFDM has been studied extensively, mainly due to the wide adoption of it in commercial communication systems \cite{Barneto2019TMTT}. In this paper, we also utilize OFDM radar for joint range-velocity estimation. Utilizing the received signal in \eqref{eq:yBSrad}, let us express the signal at the output of the digital radar combiners as
\begin{align}\label{eq:yBSrad2}
    \yBSrad = \bar{\beta}_{0,m,n} \bw_{m}^\rmH \WRF^\rmH \aBSR(\theta_0) \sum_{i=1}^{\UDL}  & \aBST^\rmH(\theta_0) \FBS \sDL \nonumber \\ & + \bw_{m}^\rmH\nULo.
\end{align}
It is important to note that the scalar term which is in product form with $\bar{\beta}_{0,m,n}$ can be constructed with the angle estimate $\hat{\theta}_0$ and the known DL signals $\sDL$. This term can be expressed as a function of $\theta$ as
\begin{equation}
    c_{m,n}(\theta) = \bw_{m}^\rmH \WRF^\rmH \aBSR(\theta) \sum_{i=1}^{\UDL}\aBST^\rmH(\theta) \FBS \sDL.
\end{equation}
We cancel the effect of this scalar and construct the image $\bZ \in \bbC^{M\times N}$ as
\begin{align}\label{eq:Z}
    [\bZ]_{m,n} &= \frac{c^{*}_{m,n}(\hat{\theta}_0)}{|c_{m,n}(\hat{\theta}_0)|^2} \yBSrad \nonumber \\
    &= \frac{c^{*}_{m,n}(\hat{\theta}_0) c_{m,n}(\theta_0)}{|c_{m,n}(\hat{\theta}_0)|^2} \bar{\beta}_{0,m,n} + \frac{c^{*}_{m,n}(\hat{\theta}_0)}{|c_{m,n}(\hat{\theta}_0)|^2} \bw_{m}^\rmH\nULo
\end{align}
If the angle estimation error is small, the constant in front of $\bar{\beta}_{0,m,n}$ would be close to 1, which is the desired case. If we investigate the image $\bZ$, we would observe that there is a progressive phase shift due to the round-trip delay on the rows, and due to the Doppler effect on the columns. Thus, we take the DFT over the rows and inverse DFT over the columns to obtain $\sfZ \in \bbC^{\tilde{M} \times \tilde{N}}$, where $\tilde{M}$ and $\tilde{N}$ are the DFT and inverse DFT lengths \cite{Barneto2019TMTT}. The obtained image has a peak which contains information related to the range and velocity of the target. Let us denote the indices of the peak location as $(\tilde{m}^\star, \tilde{n}^\star)$. The range and velocity of the target can be recovered as $r=\frac{\tilde{m}^{\star}c}{2\tilde{M}\deltaf}$ and $v=\frac{\tilde{n}^{\star}\lambda}{\tilde{N}\Ts}$, respectively.

The previous works that utilize OFDM radar often do not consider simultaneous reception of the UL signals and target reflections \cite{Barneto2019TMTT, Barneto2022TCOM, Islam2022ICC, Bayraktar2024ICC}. 
In this work, we first applied MVDR beamformer at the output of the analog combiner to suppress the UL signals to enhance the target reflections. However, there is still residual UL interference due to the limited number of RF chains at the FD BS. 
It can be shown that the residual UL interference diminishes as long as the number of subcarriers $M$ and the number of OFDM symbols $N$ are large enough \cite{Liu2023JSAC,Aditya2023OJCOMS}. The main reason is the independence of the DL and UL signals. In \eqref{eq:Z}, the multiplication of $c^{*}_{m,n}(\hat{\theta}_0)$ with the residual UL interference has the expected value of 0 due to the independence. Since the DFT and inverse DFT operations introduce summation over the interference terms, the resultant interference goes to 0 with increasing $M$ and $N$. Overall, our proposed framework has two stages, namely, MVDR beamformer and OFDM radar, that help enhancing the target reflections in the presence of UL UEs. This effect will be shown in next section.

\section{Numerical Results}
\label{sec:results}
In this section, we provide numerical results that evaluate the performance of the proposed hybrid precoder/combiner design for integrated monostatic sensing and FD MU user communication at mmWave bands.
The parameters used in simulations are given as follows.
The BS is equipped with two collocated ULAs with $\NBST = \NBSR = 64$ antennas parallel to the x-axis.
The separation between the RX and TX arrays is $ 6\lambda $ in the z-axis.
The number of RF chains at the BS is $\NBSRF=4$.
We consider two DL and two UL UEs, each of them having $\NDLs=\NULs=2$ streams.
The UEs are equipped with $\NUE=16$ antennas and $\NUERF=2$ RF chains. 
The spacing between the antenna elements for all the arrays is half wavelength.
The system operates at $28$GHz with $100$MHz bandwidth that corresponds to a noise power of $-93.8$dBm at room temperature.
The SI-to-noise ratio is set to $\rho/\varUL=80\text{dB}$.
The number of active subcarriers and OFDM symbols are set to $M=792$ and $N=\{14, 28, 56\}$, following the 5G NR standards with subcarrier spacing $\Delta f = 120$kHz and symbol duration $\Ts = 8.92\mu$s \cite{Islam2022ICC}. 

We consider $K=4$ point targets with radar cross-section of $10\textrm{m}^2$. The target angles are randomly selected from $[-60^\circ,60^\circ]$. Moreover, the range and velocity of the targets are randomly selected from $[20\textrm{m},60\textrm{m}]$ and $[10\textrm{m/s},30\textrm{m/s}]$, respectively. The UEs are randomly deployed at a distance of $50$m from the BS with the LoS angle selected from $[-60^\circ,60^\circ]$. The UE channels consist of $5$ paths. The gains of NLoS paths are $5$-$15$dB below the gain of the LoS path, while the angles are randomly selected from $[-90^\circ,90^\circ]$. The results are averaged over $100$ different realizations.

\begin{figure}[!t]
    \centering
    \includegraphics[width=0.95\linewidth]{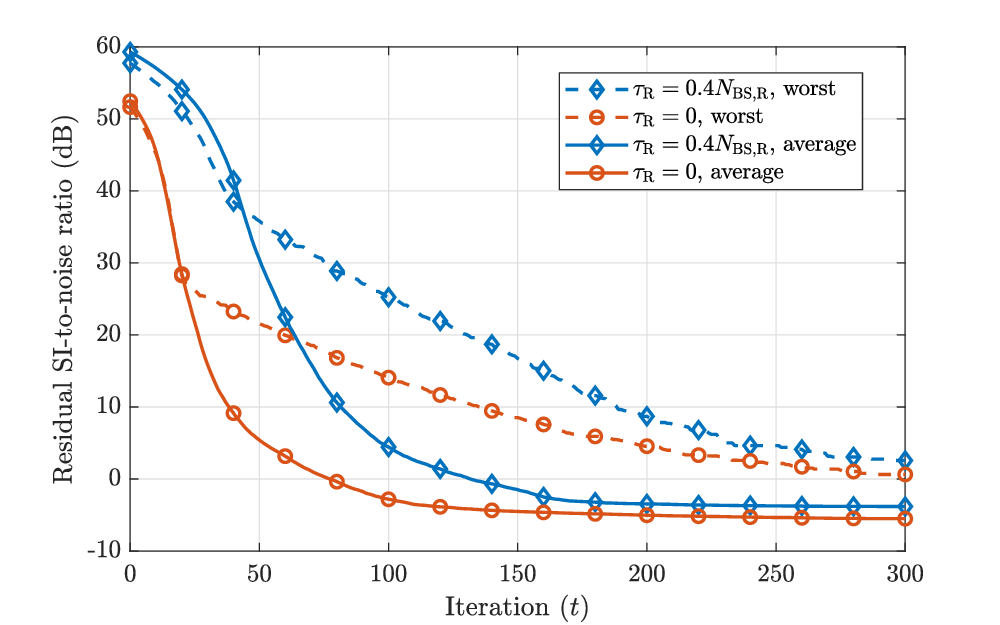}\vspace{-5pt}
    \caption{Residual SI-to-noise ratio with respect to the iterations of the analog combiner design. The dashed curves show the worst examples.}
    \label{fig:SI}
\vspace{-10pt}
\end{figure}

In the analog combiner design problem for the FD BS, the RX gain and communication gain thresholds are set to $\tauR = 0.4\NBSR$ and $\taucom = 0.5\NBSR$, respectively, and the design parameters are set to $\epsilon_1=0.3$, $\epsilon_2=0.1$, $\kappa_2=0.5$. We also consider the case without sensing requirements, which corresponds to the parameters $\tauR = 0$, $\taucom = 0.7\NBSR$ and $\kappa_2=1$. We first investigate the SI suppression performance of the joint precoder and analog combiner design for these two cases. To that end, we show the residual SI-to-noise ratio evolution with respect to the analog combiner design iterations in Fig.~\ref{fig:SI}. The iterative analog combiner design provides significant decrease in residual SI such that the resultant residual SI-to-noise ratio is below $0$dB for most cases. Another observation is that the same algorithm provides faster convergence for the case without sensing since the sensing requirement increases the angular support range of the analog combiner. In the remainder of the section, we investigate the ISAC performance of the proposed design.

\subsection{Communication Performance}
\begin{figure}[!t]
    \centering
    \includegraphics[width=0.95\linewidth]{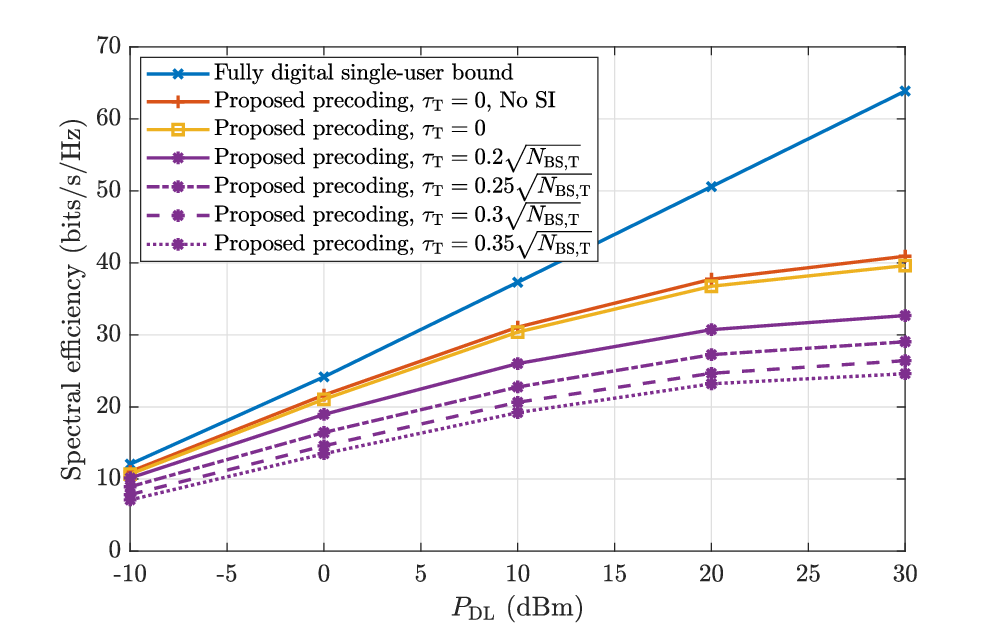}\vspace{-5pt}
    \caption{Sum spectral efficiency of DL UEs with respect to DL transmit power.}
    \label{fig:spectral_dl}
\vspace{-10pt}
\end{figure}
\begin{figure}[!t]
    \centering
    \includegraphics[width=0.95\linewidth]{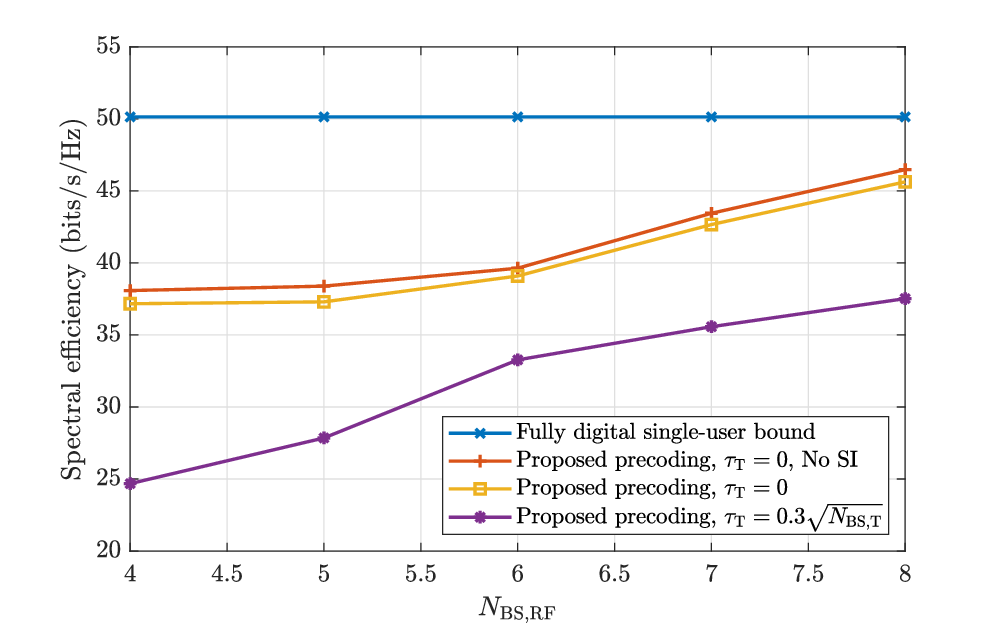}\vspace{-5pt}
    \caption{Sum spectral efficiency of DL UEs with respect to the number of RF chains at the BS. The DL transmit power is set to $\PtDL=20$dBm.}
    \label{fig:spectral_dl_RF}
\vspace{-10pt}
\end{figure}
We evaluate first the sum spectral efficiency for the DL UEs. We present three benchmarks as follows: (i) Fully digital single-user bound without interference, (ii) hybrid MU precoding without sensing and SI suppression, and (iii) hybrid MU precoding without sensing beam but with SI suppression. All the benchmarks except the first one use the variations of the proposed precoding architecture. Namely, the second benchmark is designed by omitting the SI term in the SLNR expression in \eqref{eq:SLNR2}, while the TX radar gain is set to $\tauT=0$. Similarly, the third benchmark is obtained by setting the TX radar gain to $\tauT=0$. The sum spectral efficiency results are given for the DL UEs in Fig.~\ref{fig:spectral_dl} with different TX gain thresholds. In this figure, it is observed that the hybrid factorization degrades the MU suppression, even without the integration of sensing. Hence, we observe saturation behavior at high transmit power. Introduction of sensing further degrades the performance as expected, which is the result of increase in MU interference for the DL UEs. To decrease the degradation, it is reasonable to increase the number of RF chains at the BS above the number of total steams, i.e., $\NBSRF>\UDL\NDLs$. In Fig.~\ref{fig:spectral_dl_RF}, we fix the DL transmit power to $\PtDL=20$dBm, and evaluate the sum spectral efficiency with respect to the number of RF chains at the BS. It can be seen that increasing the number of RF chains significantly improves the performance with or without sensing.

\begin{figure}[!t]
    \centering
    \includegraphics[width=0.95\linewidth]{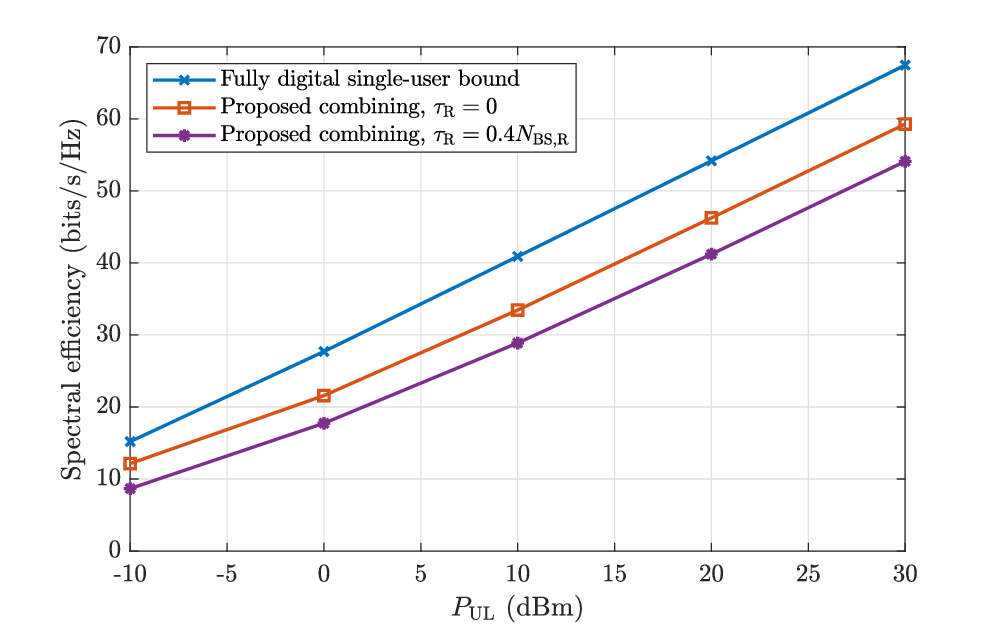}\vspace{-5pt}
    \caption{Sum spectral efficiency of UL UEs with respect to UL transmit power.}
    \label{fig:spectral_ul}
\vspace{-5pt}    
\end{figure}

We also evaluate the sum spectral efficiency for UL UEs when the transmit power of the DL transmission and the TX radar gain are fixed to $\PtDL=20\text{dBm}$, $\tauT=0.3\sqrt{\NBST}$, respectively. The same transmit power $P_{\rm UL}$ is used for all UL UEs. We utilize the following benchmarks: (i) Fully digital single-user bound without interference, and (ii) hybrid MU precoding without sensing. The sum spectral efficiency results are given for the UL UEs in Fig.~\ref{fig:spectral_ul}. The results show that the proposed architecture works remarkably well. Since the MU separation is handled at the digital domain, we do not observe saturation. Furthermore, the loss caused by the integration of sensing is lower compared to the loss observed in DL communication.

\subsection{Sensing Performance}
\begin{figure}[!t]
    \centering
    \includegraphics[width=0.9\linewidth]{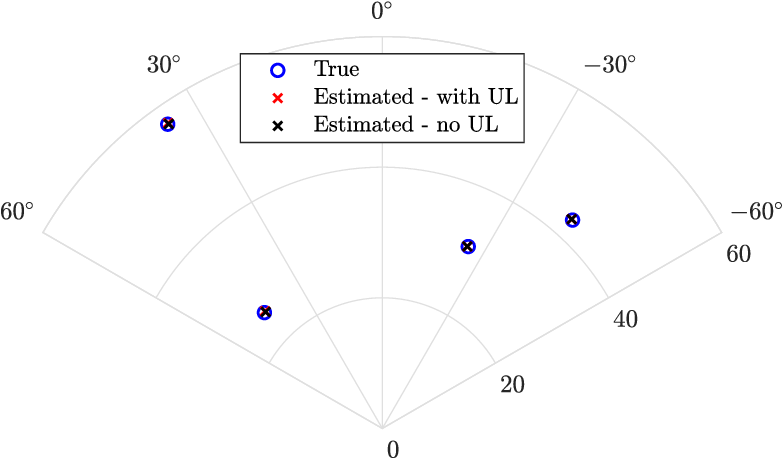}
    \caption{Angle-range map of the targets.}
    \label{fig:angle-range}
    \vspace{-10pt}
\end{figure}
To evaluate the sensing performance, we employ the same system parameters. The results are obtained with $\PtDL=20\text{dBm}$, $\PtUL=10\text{dBm}$ and $\tauT=0.25\sqrt{\NBST}$. This is a reasonable setting since UL UEs operate with lower power levels. The initial angle estimates are generated around the true value with an error modeled as Gaussian random variable with variance $1^\circ$. The data streams utilize QPSK symbols. We show the estimation results with and without the UL signals. In Fig.~\ref{fig:angle-range}, we show an angle-range map for $N=14$ OFDM symbols. It can be seen that the angle and range estimates are very closely matched with the true positions, and the presence of UL signals does not have an effect. The reasons are threefold: (i) We have an initial angle estimate, and the angle refinement stage discards the contributions of UL signals which lead to accurate angle estimates, (ii) the MVDR beamformers suppress the UL signals, and (iii) the number of subcarriers $M$ is large enough to eliminate the effect of the residual UL signals in OFDM radar processing.

\begin{figure}[!t]
\centering
\subfloat[$N=14$]{\includegraphics[width=0.9\linewidth]{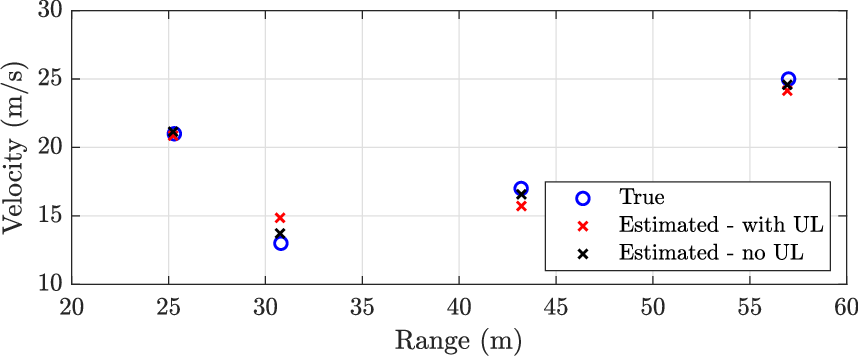}}
\vspace{-1pt}
\subfloat[$N=28$]{\includegraphics[width=0.9\linewidth]{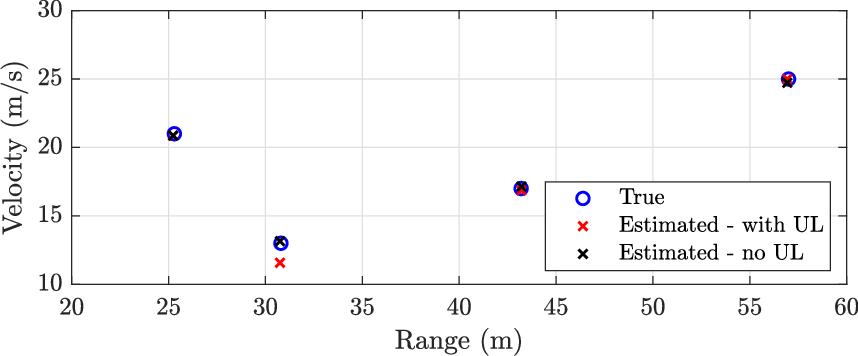}}
\vspace{-1pt}
\subfloat[$N=56$]{\includegraphics[width=0.9\linewidth]{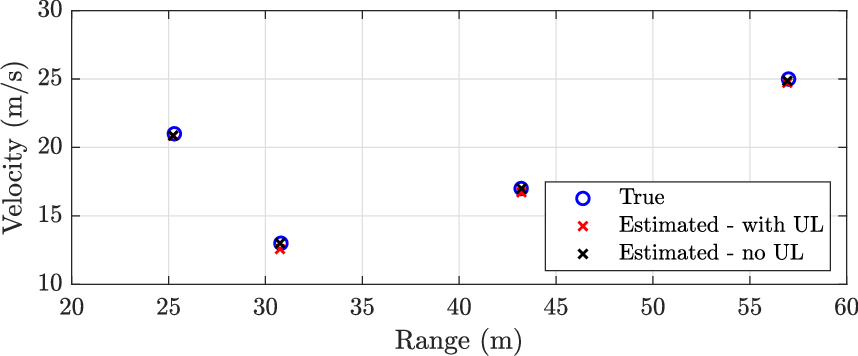}}
\caption{Range-velocity maps of the targets for different number of OFDM symbols.}
\vspace{-10pt}
\label{fig:range-velocity}
\end{figure}

For the range-velocity maps, which are given in Fig.~\ref{fig:range-velocity}, we use $N=\{14, 28, 56\}$ OFDM symbols. We observe erroneous velocity estimates when $N=14$, which is expected, even for the case without UL signals. It is important to note that it is still possible to have reliable estimates in the presence of UL signals even though the number of OFDM symbols is very limited, thanks to the MVDR beamformers which suppress the UL signals. If the number of OFDM symbols is increased, we observe significantly better velocity estimates. It can be seen that the velocity estimation error is very small when we have $N=56$ OFDM symbols with or without UL signals. These results show that it is possible to estimate target parameters as long as the number of subcarriers and the number of OFDM symbols are large enough. However, large number of OFDM symbols may create practical issues. Since the channel coherence time is limited, the channel estimates need to be updated, thus, it may not be possible to collect a large number consecutive OFDM symbols. Alternative solutions to circumvent this issue are to: (i) either increase the TX/RX radar gain, (ii) decrease the transmit power of UL users, or (iii) increase the number of RF chains beyond to increase the degrees-of-freedom of MVDR beamformer to suppress the UL signals more efficiently. In those cases, more reliable estimates can be obtained even with a small number of OFDM symbols.

\section{Conclusion}
In this paper, we proposed the first SI-aware hybrid precoding and combining architecture for integrated monostatic sensing and FD MU communication at mmWave bands and its associated processing at the dual receiver. This is the first algorithmic solution that enables simultaneous sensing, downlink and uplink communication accounting for SI. To solve this challenging problem, we utilized SLNR-based precoding and an iterative convex relaxation-based analog combining solution to jointly suppress the SI while simultaneously satisfying sensing and communication requirements at the FD BS. Since the received signal at the output of the analog combiner contains both UL signals and target reflections, we developed a novel interference-aware digital combining and parameter estimation framework. The simulation results showed the robustness of the proposed design to strong SI while providing simultaneous sensing and communication. Additionally, we showed that joint sensing and UL communication can be enabled for FD ISAC systems operating at mmWave bands.

\bibliographystyle{IEEEtran}
\bibliography{refs}

\begin{thebibliography}{10}
\providecommand{\url}[1]{#1}
\csname url@samestyle\endcsname
\providecommand{\newblock}{\relax}
\providecommand{\bibinfo}[2]{#2}
\providecommand{\BIBentrySTDinterwordspacing}{\spaceskip=0pt\relax}
\providecommand{\BIBentryALTinterwordstretchfactor}{4}
\providecommand{\BIBentryALTinterwordspacing}{\spaceskip=\fontdimen2\font plus
\BIBentryALTinterwordstretchfactor\fontdimen3\font minus
  \fontdimen4\font\relax}
\providecommand{\BIBforeignlanguage}[2]{{%
\expandafter\ifx\csname l@#1\endcsname\relax
\typeout{** WARNING: IEEEtran.bst: No hyphenation pattern has been}%
\typeout{** loaded for the language `#1'. Using the pattern for}%
\typeout{** the default language instead.}%
\else
\language=\csname l@#1\endcsname
\fi
#2}}
\providecommand{\BIBdecl}{\relax}
\BIBdecl

\bibitem{Liu2022JSAC}
F.~Liu, Y.~Cui, C.~Masouros, J.~Xu, T.~X. Han, Y.~C. Eldar, and S.~Buzzi,
  ``Integrated sensing and communications: Toward dual-functional wireless
  networks for {6G} and beyond,'' \emph{IEEE J. Sel. Areas Commun.}, vol.~40,
  no.~6, pp. 1728--1767, 2022.

\bibitem{Proceedings2024NGP}
N.~Gonz\'{a}lez-Prelcic, M.~F. Keskin, O.~Kaltiokallio, M.~Valkama, D.~Dardari,
  X.~Shen, Y.~Shen, M.~Bayraktar, and H.~Wymeersch, ``The integrated sensing
  and communication revolution for {6G}: Vision, techniques, and
  applications,'' \emph{Proc. IEEE}, 2024.

\bibitem{Barneto2019TMTT}
C.~Baquero~Barneto, T.~Riihonen, M.~Turunen, L.~Anttila, M.~Fleischer,
  K.~Stadius, J.~Ryynänen, and M.~Valkama, ``Full-duplex {OFDM} radar with
  {LTE} and {5G} {NR} waveforms: Challenges, solutions, and measurements,''
  \emph{IEEE Trans. Microw. Theory Tech.}, vol.~67, no.~10, pp. 4042--4054,
  2019.

\bibitem{Heath2016JSTSP}
R.~W. Heath, N.~González-Prelcic, S.~Rangan, W.~Roh, and A.~M. Sayeed, ``An
  overview of signal processing techniques for millimeter wave {MIMO}
  systems,'' \emph{IEEE J. Sel. Topics Signal Process.}, vol.~10, no.~3, pp.
  436--453, 2016.

\bibitem{Barneto2021WC}
C.~B. Barneto, S.~D. Liyanaarachchi, M.~Heino, T.~Riihonen, and M.~Valkama,
  ``Full duplex radio/radar technology: The enabler for advanced joint
  communication and sensing,'' \emph{IEEE Wireless Commun.}, vol.~28, no.~1,
  pp. 82--88, 2021.

\bibitem{Smida2024Proc}
B.~Smida, R.~Wichman, K.~E. Kolodziej, H.~A. Suraweera, T.~Riihonen, and
  A.~Sabharwal, ``In-band full-duplex: The physical layer,'' \emph{Proc. IEEE},
  pp. 1--30, 2024.

\bibitem{Keskin2021TSP}
M.~F. Keskin, V.~Koivunen, and H.~Wymeersch, ``Limited feedforward waveform
  design for {OFDM} dual-functional radar-communications,'' \emph{IEEE Trans.
  Signal Process.}, vol.~69, pp. 2955--2970, 2021.

\bibitem{Jiang2022SystJ}
Z.-M. Jiang, M.~Rihan, P.~Zhang, L.~Huang, Q.~Deng, J.~Zhang, and E.~M.
  Mohamed, ``Intelligent reflecting surface aided dual-function radar and
  communication system,'' \emph{IEEE Syst. J.}, vol.~16, no.~1, pp. 475--486,
  2022.

\bibitem{Bazzi2023TWC}
A.~Bazzi and M.~Chafii, ``On outage-based beamforming design for
  dual-functional radar-communication {6G} systems,'' \emph{IEEE Trans.
  Wireless Commun.}, vol.~22, no.~8, pp. 5598--5612, 2023.

\bibitem{Roberts2021WC}
I.~P. Roberts, J.~G. Andrews, H.~B. Jain, and S.~Vishwanath, ``Millimeter-wave
  full duplex radios: New challenges and techniques,'' \emph{IEEE Wireless
  Commun.}, vol.~28, no.~1, pp. 36--43, 2021.

\bibitem{Sabharwal2014JSAC}
A.~Sabharwal, P.~Schniter, D.~Guo, D.~W. Bliss, S.~Rangarajan, and R.~Wichman,
  ``In-band full-duplex wireless: Challenges and opportunities,'' \emph{IEEE J.
  Sel. Areas Commun.}, vol.~32, no.~9, pp. 1637--1652, 2014.

\bibitem{Valcarce2019FD}
R.~López-Valcarce and N.~González-Prelcic, ``Analog beamforming for
  full-duplex millimeter wave communication,'' in \emph{Proc. 16th Int. Symp.
  Wireless Commun. Syst. (ISWCS)}, 2019, pp. 687--691.

\bibitem{Satyanarayana2019TVT}
K.~Satyanarayana, M.~El-Hajjar, P.-H. Kuo, A.~Mourad, and L.~Hanzo, ``Hybrid
  beamforming design for full-duplex millimeter wave communication,''
  \emph{IEEE Trans. Veh. Technol.}, vol.~68, no.~2, pp. 1394--1404, 2019.

\bibitem{Palacios2019GLOBECOM}
J.~Palacios, J.~Rodriguez-Fernandez, and N.~Gonzalez-Prelcic, ``Hybrid
  precoding and combining for full-duplex millimeter wave communication,'' in
  \emph{Proc. IEEE Global Commun. Conf. (GLOBECOM)}, 2019, pp. 1--6.

\bibitem{Valcarce2020FD}
R.~López-Valcarce and M.~Martínez-Cotelo, ``Full-duplex {mmWave}
  communication with hybrid precoding and combining,'' in \emph{Proc. 28th Eur.
  Signal Process. Conf. (EUSIPCO)}, 2021, pp. 1752--1756.

\bibitem{Roberts2021TWC}
I.~P. Roberts, J.~G. Andrews, and S.~Vishwanath, ``Hybrid beamforming for
  millimeter wave full-duplex under limited receive dynamic range,'' \emph{IEEE
  Trans. Wireless Commun.}, vol.~20, no.~12, pp. 7758--7772, 2021.

\bibitem{Koc2021OJCOMS}
A.~Koc and T.~Le-Ngoc, ``Full-duplex {mmWave} massive {MIMO} systems: A joint
  hybrid precoding/combining and self-interference cancellation design,''
  \emph{IEEE Open J. Commun. Soc.}, vol.~2, pp. 754--774, 2021.

\bibitem{Sheemar2022OJCOMS}
C.~K. Sheemar, C.~K. Thomas, and D.~Slock, ``Practical hybrid beamforming for
  millimeter wave massive {MIMO} full duplex with limited dynamic range,''
  \emph{IEEE Open J. Commun. Soc.}, vol.~3, pp. 127--143, 2022.

\bibitem{Valcarce2022TWC}
R.~López-Valcarce and M.~Martínez-Cotelo, ``Full-duplex {mmWave} {MIMO} with
  finite-resolution phase shifters,'' \emph{IEEE Trans. Wireless Commun.},
  vol.~21, no.~11, pp. 8979--8994, 2022.

\bibitem{Liyanaarachchi2021JCS}
S.~D. Liyanaarachchi, C.~B. Barneto, T.~Riihonen, M.~Heino, and M.~Valkama,
  ``Joint multi-user communication and {MIMO} radar through full-duplex hybrid
  beamforming,'' in \emph{Proc. 1st IEEE Int. Online Symp. Joint Commun.
  Sensing (JCS)}, 2021, pp. 1--5.

\bibitem{Barneto2022TCOM}
C.~B. Barneto, T.~Riihonen, S.~D. Liyanaarachchi, M.~Heino,
  N.~González-Prelcic, and M.~Valkama, ``Beamformer design and optimization
  for joint communication and full-duplex sensing at mm-{Waves},'' \emph{IEEE
  Trans. Commun.}, vol.~70, no.~12, pp. 8298--8312, 2022.

\bibitem{Islam2022ICC}
M.~A. Islam, G.~C. Alexandropoulos, and B.~Smida, ``Integrated sensing and
  communication with millimeter wave full duplex hybrid beamforming,'' in
  \emph{Proc. IEEE Int. Conf. Commun. (ICC)}, 2022, pp. 4673--4678.

\bibitem{Islam2022GLOBECOM}
M.~A. Islam, G.~C. Alexandropoulos, and B.~Smida, ``Simultaneous multi-user
  {MIMO} communications and multi-target tracking with full duplex radios,'' in
  \emph{Proc. IEEE Globecom Workshops (GC Wkshps)}, 2022, pp. 19--24.

\bibitem{Bayraktar2023CAMSAP}
M.~Bayraktar, C.~Rusu, N.~González-Prelcic, and H.~Chen, ``Self-interference
  aware codebook design for full-duplex joint sensing and communication systems
  at {mmWave},'' in \emph{Proc. IEEE 9th Int. Workshop Comput. Adv.
  Multi-Sensor Adaptive Process. (CAMSAP)}, 2023, pp. 231--235.

\bibitem{Bayraktar2024ICC}
M.~Bayraktar, N.~González-Prelcic, and H.~Chen, ``Hybrid precoding and
  combining for {mmWave} full-duplex joint radar and communication systems
  under self-interference,'' in \emph{Proc. IEEE Int. Conf. Commun. (ICC)},
  2024, pp. 1--6.

\bibitem{Liu2023JSAC}
Z.~Liu, S.~Aditya, H.~Li, and B.~Clerckx, ``Joint transmit and receive
  beamforming design in full-duplex integrated sensing and communications,''
  \emph{IEEE J. Sel. Areas Commun.}, vol.~41, no.~9, pp. 2907--2919, 2023.

\bibitem{He2023JSAC}
Z.~He, W.~Xu, H.~Shen, D.~W.~K. Ng, Y.~C. Eldar, and X.~You, ``Full-duplex
  communication for {ISAC}: Joint beamforming and power optimization,''
  \emph{IEEE J. Sel. Areas Commun.}, vol.~41, no.~9, pp. 2920--2936, 2023.

\bibitem{Talha2023Asilomar}
M.~Talha, B.~Smida, M.~A. Islam, and G.~C. Alexandropoulos, ``Multi-target
  two-way integrated sensing and communications with full duplex {MIMO}
  radios,'' in \emph{Proc. 57th Asilomar Conf. Signals, Syst., Comput.}, 2023,
  pp. 1661--1667.

\bibitem{Elbir2023SPM}
A.~M. Elbir, K.~V. Mishra, S.~A. Vorobyov, and R.~W. Heath, ``Twenty-five years
  of advances in beamforming: From convex and nonconvex optimization to
  learning techniques,'' \emph{IEEE Signal Process. Mag.}, vol.~40, no.~4, pp.
  118--131, 2023.

\bibitem{Sadek2007TWC}
M.~Sadek, A.~Tarighat, and A.~H. Sayed, ``A leakage-based precoding scheme for
  downlink multi-user {MIMO} channels,'' \emph{IEEE Trans. Wireless Commun.},
  vol.~6, no.~5, pp. 1711--1721, 2007.

\bibitem{Ghojogh2019arXiv}
B.~Ghojogh, F.~Karray, and M.~Crowley, ``Eigenvalue and generalized eigenvalue
  problems: Tutorial,'' \emph{arXiv preprint arXiv:1903.11240}, 2019.

\bibitem{Coma2018JSTSP}
J.~P. González-Coma, J.~Rodríguez-Fernández, N.~González-Prelcic,
  L.~Castedo, and R.~W. Heath, ``Channel estimation and hybrid precoding for
  frequency selective multiuser {mmWave} {MIMO} systems,'' \emph{IEEE J. Sel.
  Topics Signal Process.}, vol.~12, no.~2, pp. 353--367, 2018.

\bibitem{Fernandez2018SPAWC}
J.~Rodriguez-Fernández and N.~Gonzálcz-Prelcic, ``Low-complexity multiuser
  hybrid precoding and combining for frequency selective millimeter wave
  systems,'' in \emph{Proc. IEEE 19th Int. Workshop Signal Process. Advances
  Wireless Commun. (SPAWC)}, 2018, pp. 1--5.

\bibitem{Zhang2019IET}
D.~Zhang, Y.~Wang, X.~Li, and W.~Xiang, ``Hybrid beamforming for downlink
  multiuser millimetre wave {MIMO}-{OFDM} systems,'' \emph{IET Commun.},
  vol.~13, no.~11, pp. 1557--1564, 2019.

\bibitem{Nesterov2012SIAM}
Y.~Nesterov, ``Efficiency of coordinate descent methods on huge-scale
  optimization problems,'' \emph{SIAM J. Optim.}, vol.~22, no.~2, pp. 341--362,
  2012.

\bibitem{Grant2014cvx}
M.~Grant and S.~Boyd, ``{CVX}: Matlab software for disciplined convex
  programming, version 2.1,'' 2014.

\bibitem{VanTrees2002}
H.~L. Van~Trees, \emph{Optimum array processing: Part {IV} of detection,
  estimation, and modulation theory}.\hskip 1em plus 0.5em minus 0.4em\relax
  John Wiley \& Sons, 2002.

\bibitem{Aditya2023OJCOMS}
S.~Aditya, O.~Dizdar, B.~Clerckx, and X.~Li, ``Sensing using coded
  communications signals,'' \emph{IEEE Open J. Commun. Soc.}, vol.~4, pp.
  134--152, 2023.

\end{thebibliography}

\end{document}